\documentclass[a4paper,12pt]{article}
\pdfoutput=1

\usepackage{amssymb,amsmath,bm}
\usepackage{a4wide}
\usepackage{color}
\usepackage{slashed}
\usepackage{graphicx}
\usepackage{amsfonts}
\usepackage{lscape}
\usepackage{hyperref}
\usepackage{amsthm}
\hypersetup{
    colorlinks=true, 
    linktoc=all,     
    linkcolor=blue,  
    citecolor= blue
}
\usepackage{booktabs}
\usepackage{enumerate}
\usepackage{array}
\usepackage{rotating}
\usepackage[numbers, sort&compress]{natbib}
\usepackage{float}
\usepackage[utf8]{inputenc}
\usepackage[T1]{fontenc}
\newcommand{\ord}[1]{\mathcal{O}\left( #1 \right)}

\newcommand{\be}{\begin{equation}}
\newcommand{\ee}{\end{equation}}

\newcommand{\eref}[1]{Eq.~(\ref{#1})}

\newcommand{\beq}{\begin{equation}} 
\newcommand{\eeq}{\end{equation}} 
\newcommand{\ba}{\begin{array}}  
\newcommand{\ea}{\end{array}} 
\newcommand{\bea}{\begin{eqnarray}}  
\newcommand{\eea}{\end{eqnarray} }  
\newcommand{\bal}{\begin{align}}
\newcommand{\eal}{\end{align}}   
\newcommand{\bi}{\begin{itemize}}  
\newcommand{\ei}{\end{itemize}}  
\newcommand{\ben}{\begin{enumerate}}  
\newcommand{\een}{\end{enumerate}}  
\newcommand{\bc}{\begin{center}}
\newcommand{\ec}{\end{center}} 
\newcommand{\bt}{\begin{table}}
\newcommand{\et}{\end{table}}  
\newcommand{\btb}{\begin{tabular}}
\newcommand{\etb}{\end{tabular}}

\newcommand{\MET}{E\llap{/\kern1.5pt}_T}

\definecolor{Grn}{rgb}{0.,0.6,0.}

\newcommand{\Red}[1]{{\textcolor{red}{#1}}}

\begin{document}

\begin{titlepage}
\vspace*{-1.0truecm}
\begin{flushright}
ULB-TH/18-07\\
 \vspace*{2mm}
 \end{flushright}
\vspace{0.8truecm}

\begin{center}
\boldmath

\textbf{
{\LARGE  Singlet-Doublet Dark Matter Freeze-in:}  \\[4mm] {\Large  LHC displaced signatures versus cosmology}
}
\unboldmath
\end{center}

\vspace{0.4truecm}

\begin{center}
{\large
{\bf Lorenzo  Calibbi$\,^a$,  Laura Lopez-Honorez$\,^{b,c}$, 
\\[1.5mm] Steven Lowette$\,^d$, and Alberto Mariotti$\,^{c,d}$}
}
\vspace{0.6truecm}

{\footnotesize

$^a${\sl CAS Key Laboratory of Theoretical Physics, Institute of Theoretical Physics, \\ Chinese Academy of Sciences, Beijing 100190, P.~R.~China \vspace{0.2truecm}}

$^b${\sl Service de Physique Th\'eorique, Universit\'e Libre de Bruxelles, C.P. 225,  B-1050 Brussels, Belgium \vspace{0.2truecm}}

$^c${\sl Theoretische Natuurkunde, Vrije Universiteit Brussel, Pleinlaan 2,   B-1050 Brussels, Belgium \vspace{0.2truecm}}

$^d${\sl Inter-University Institute for High Energies, Vrije Universiteit Brussel, \\ Pleinlaan 2,  B-1050 Brussels, Belgium}

}
\end{center}

\vspace{0.4truecm}
\begin{abstract}
\noindent 
We study the Singlet-Doublet dark matter model in the
  regime of feeble couplings, where the dark matter abundance is
  obtained via the freeze-in mechanism.  As a consequence of the small
  couplings, the heavier particles in the model are 
  long-lived with decay length at typical scales of collider experiments.
    We analyse the collider signatures of the
  model, characterised by displaced $h$ and $Z$ bosons plus missing
  momentum, employing current LHC searches for displaced vertices and
  missing energy to significantly constrain the parameter space of the
  model.  We also take into account the cosmological bounds
    relevant for our light dark matter candidate arising
    from Lyman-$\alpha$ forest constraints. Our analysis emphasises
    the interplay between displaced signatures at the LHC and cosmology for
    dark matter candidates whose relic abundance is obtained through
    the freeze-in mechanism.
 \end{abstract}

\end{titlepage}
\tableofcontents




\section{Introduction}
\setcounter{equation}{0}

Compelling observational evidence supports the existence of dark
matter (DM), the most abundant form of matter in the
Universe~\cite{Ade:2015xua}. Yet, at present, the nature of the dark
sector, possibly including dark matter particle(s) and new
mediators driving the interactions between dark matter and the
Standard Model (SM), is still unknown. In the last decades most of the
attention has been devoted to the weakly interacting massive particle
(WIMP) paradigm, in which dark matter is a new type of elementary
particle with weak-type interactions with the SM.  In WIMP models the
abundance of the dark matter is obtained through the freeze-out
mechanism.
The weak couplings involved typically imply a viable\,---\,i.e.~giving rise to the 
observed relic abundance\,---\,dark matter particle
with mass of the order of the electroweak (EW) scale. 
This remarkable coincidence\,---\,the so-called \emph{WIMP miracle}\,---\,has motivated a large effort 
in the study of WIMP dark matter scenarios both in top-down approaches, triggered by
supersymmetry, and bottom-up approaches focusing on effective theories
and simplified models. In the latter framework, an important effort
has been deployed in the last years to characterise dark matter
simplified model parameter space, to search for the dark matter
particle and the associated mediators at colliders, and to explore the
complementarity between the LHC, the direct and indirect detection
experiments, see e.g.~\cite{Abdallah:2015ter,Abercrombie:2015wmb,Albert:2017onk}. 
The recurrent null results in the search for WIMPs
(both at the LHC and in direct/indirect detection experiments) provide
good motivations though to take a step back and to investigate
alternative dark matter paradigms and to study thoroughly their
phenomenology.

Here we focus on the so-called \emph{freeze-in} mechanism for
producing a \emph{feebly} coupled dark matter candidate,
i.e.~particles that were not in thermal equilibrium with the SM in the
early universe, see e.g.~\cite{Hall:2009bx,Chu:2011be,Bernal:2017kxu}.
Because of the very small coupling involved, these models can give
rise to displaced signatures at the LHC in terms of long-lived
mediators decaying into dark matter plus SM fields, as it has already
been underlined in a number of
works~\cite{Hall:2009bx,Co:2015pka,Hessler:2016kwm,DEramo:2017ecx,Brooijmans:2018xbu}.
Long-lived particles and displaced signatures at colliders in relation
to dark matter simplified models have been discussed also in
\cite{Chang:2009sv,Davoli:2017swj,Buchmueller:2017uqu,Garny:2017rxs,Ghosh:2017vhe,Garny:2018icg,Davoli:2018mau}.
The requirement of a viable frozen-in dark matter scenario giving rise
to displacement at colliders of detector size points directly towards
light dark matter candidates, with a mass of the order of the
keV~\cite{Hall:2009bx}. This is typically the mass scale currently
tested by cosmology and astrophysics probes in the framework of warm
dark matter scenarios, see
e.g.~\cite{Viel:2013apy,Yeche:2017upn,Irsic:2017ixq}, that lead to the
suppression of the small scale structure formation and, by the same
token, can help to alleviate the small scale
crisis~\cite{Bullock:2017xww,Klypin:1999uc, Moore:1999nt,
  BoylanKolchin:2011dk, Moore:1999gc, Springel:2008cc} in the
$\Lambda$CDM (the Standard Cosmological scenario) see e.g.~the
discussion in~\cite{Bode:2000gq, Zavala:2009ms, Lovell:2011rd,Schneider:2011yu, Lovell:2013ola, Kennedy:2013uta, Lovell:2016nkp,
  Lopez-Honorez:2017csg}. On the other hand, due to the feeble
coupling involved, direct and indirect detection dark matter searches
are challenging, see however
e.g.~\cite{Hochberg:2017wce,Knapen:2017ekk,Heikinheimo:2018duk,Bernal:2018ins}.

In this paper we
will study how the combination of collider and cosmological
constraints can probe a significant portion of
the parameter space of frozen-in dark matter
models. 
First, note that the long-lived mediators typically possess sizeable couplings with the SM particles that keep them in thermal equilibrium 
with the SM bath, and thus they can copiously be produced at the LHC.
Second,
the reach of displaced vertex signatures at the LHC can actually
extend beyond the regime where the mean decay length of the
heavy mediator, $c\tau$, is within the tracking detector with typical radius $\simeq~1$ m.
This is because of the background-free nature of displaced signatures
and because of the exponential decay distribution.\footnote{The latter is
indeed such that a substantial fraction of the mediators produced at
LHC would decay within the detector even for $c\tau>1$ m, see
e.g.~\cite{Ishiwata:2008tp}.}
In general, the study of displaced
signatures with missing energy at the LHC is thus a powerful probe of
frozen-in dark matter in the small mass regime (keV to a few MeV),
complementing the existing constraints from astro-physics and
cosmology on light free-streeming DM candidates.

For concreteness, we focus here on the Singlet-Doublet dark matter
model~\cite{Mahbubani:2005pt,DEramo:2007anh,Enberg:2007rp,Cohen:2011ec,Cheung:2013dua,Abe:2014gua,Calibbi:2015nha,Freitas:2015hsa,Egana-Ugrinovic:2017jib,Lopez-Honorez:2017ora,Esch:2018ccs,Arcadi:2018pfo}.
In the Singlet-Doublet model, the SM is augmented with a pair of
electroweak doublet Weyl fermions and one Majorana singlet,
interacting with the SM Higgs via Yukawa couplings.  The dark sector
contains thus a charged fermion and three neutral fermions, whose lightest
state constitutes a stable dark matter candidate.  Within the
freeze-in regime of such model, the Yukawa couplings are feeble and
thus the dark matter particle is mostly singlet while the role of the
long-lived mediators is played by the components of the electroweak
doublets.
Two of them are neutral and hence a smoking-gun signature of this
scenario at the LHC consists of displaced Higgs bosons or $Z$-bosons
plus missing momentum.  
We analyse the LHC sensitivity on such interesting final states 
by recasting an existing ATLAS search for displaced vertices and missing energy \cite{Aaboud:2017iio}.
Making use of the large statistics already collected,
we show that the proper decay length that can
be (will be) tested is actually significantly larger than the detector
size, reaching more than 10 (100) meters. In the corresponding parts
of the parameter space with the correct dark matter abundance, this
currently constrains the dark matter mass to be as large as 500 keV and
could reach a few MeV with 300 fb$^{-1}$, hence well beyond the warm
dark matter regime.

The rest of the paper is organised as follows. In Section \ref{sec:model}, we
discuss the model in the feeble-coupling regime and analyse the decay
modes of the dark sector particles. Section \ref{sec:FI} is devoted to the
calculation of the abundance of dark matter produced through the freeze-in mechanism.
We then
study the constraints on light dark matter from cosmology in Section \ref{sec:cosmo-bounds}.  
In Section \ref{sec:LHC}, we analyse in detail the main collider signatures\,---\,disappearing charged tracks, as well as displaced Higgs or $Z$ plus missing transverse momentum ($\MET$)\,---\,of our scenario and we present the recasting of the ATLAS search \cite{Aaboud:2017iio}.
In Section \ref{sec:discussion}, we combine the results of the previous sections
and show the interplay between collider and cosmological signatures in
probing our model of freeze-in dark matter.
We summarise and conclude in Section \ref{sec:conclusions}, while 
we present some technical details in the Appendices.

\section{The feebly-coupled Singlet-Doublet DM model}
\label{sec:model}

We perform our analysis within the Singlet-Doublet dark matter model
\cite{Mahbubani:2005pt}, which consists in adding to the SM a pair of Weyl doublet fermions, $\psi_u$
and $\psi_d$, with opposite hypercharges and one fermionic singlet,
$\psi_s$:
\begin{equation} (\psi_u)_{2,\,\frac{1}{2}} = \left(
\begin{array}{c}
\psi^{+} \\
\psi_u^0
\end{array} 
\right),
\qquad
(\psi_d)_{2,\,-\frac{1}{2}} = 
\left(
\begin{array}{c}
\psi_d^0 \\
\psi^{-}
\end{array} 
\right),
\qquad
(\psi_s)_{1,\,0}\,.
\end{equation}
The subscripts indicate the $SU(2)_L\times U(1)_Y$ quantum numbers and
these new fields are assumed to be odd under an unbroken $Z_2$
symmetry, under which the SM fields are even, so as to guarantee the
stability of dark matter. 

The mass terms and Yukawa interactions of the model read
\begin{equation}
\label{eq:lagr}
-\mathcal{L} \supset \mu ~\psi_d \cdot \psi_u + y_d ~\psi_d \cdot H \,\psi_s + y_u  ~H^{\dagger} \psi_u \,\psi_s + \frac{1}{2} m_s ~\psi_s \psi_s +h.c.\,,
\end{equation}
where $H$ is the Higgs doublet (with hypercharge $1/2$) and $\cdot$
indicates a contraction of the $SU(2)_L$ indices through the
antisymmetric tensor $\epsilon_{ab}$, see appendix~\ref{app:masses} for more details.  
For later convenience, we also define the following alternative
parameterisation of the two Yukawa couplings:
\begin{equation}
y_u \equiv y \sin\theta,\quad y_d \equiv y \cos\theta.
\end{equation}
As is apparent from the field content, the model is a generalisation
of the ``Bino-Higgsino'' system of supersymmetric models with free
couplings $y_u$ and $y_d$ (whereas supersymmetry would relate them to
the SM gauge couplings).

\subsection{The spectrum}

Upon EW-symmetry breaking, the Lagrangian in \eref{eq:lagr} leads to mixing among 
the neutral components $(\psi_s,\,\psi^0_d,\,\psi^0_u )$ of the $Z_2$-odd fermions.
The resulting mass matrix reads
\begin{equation}
\mathcal{M} = 
\left(
\begin{array}{ccc}
m_s & \frac{y_d v}{\sqrt{2}} & \frac{y_u v}{\sqrt{2}} \\
 \frac{y_d v}{\sqrt{2}}  & 0& \mu \\
\frac{y_u v}{\sqrt{2}} & \mu &0
\end{array}
\right),
\end{equation}
where $v\simeq 246$ GeV is the vev of the Higgs field.
The above matrix is diagonalised by a rotation matrix $U$, $U  \mathcal{M}  U^{T} = \hat{\mathcal{M}}$. 
The mass eigenstates are then given by
\begin{equation}
\label{eq:Udef}
\left(  \chi_1, \chi_2,  \chi_3\right)^T= U \, \left(  \psi_s, \psi_d^0,  \chi_u^0\right)^T.
\end{equation}
We employ the convention $|m_{\chi_1}|<|m_{\chi_2}|<|m_{\chi_3}|$, thus our dark matter candidate is $\chi_1$.

The model has already been extensively investigated within the
framework of the freeze-out mechanism of dark matter
production~\cite{Mahbubani:2005pt,DEramo:2007anh,Enberg:2007rp,Cohen:2011ec,Cheung:2013dua,Abe:2014gua,Calibbi:2015nha,Freitas:2015hsa,Egana-Ugrinovic:2017jib,Lopez-Honorez:2017ora,Esch:2018ccs,Arcadi:2018pfo}. The
latter requires the couplings $y_u$ and $y_d$ to be of the order of
$10^{-2}-1$ for the Yukawa interactions to drive the relic abundance
to the observed one. In that case the model is constrained by direct
and indirect detection experiments and also features interesting
collider signatures, typically resembling the Bino-Higgsino system of
supersymmetric models (but with arbitrary Yukawa couplings). In
contrast, in this work, we focus on the freeze-in mechanism of dark
matter production associated to very feeble Singlet-Doublet
interactions. As it will become clear from the discussion in
Section~\ref{sec:FI}, the typical Yukawa couplings of interest for our
analysis range from $10^{-9}$ to $10^{-6}$ and a large mass difference
between the singlet and the doublet mass-scales will have to be
considered: $|m_s| \ll |\mu|$. As a result, the model features
suppressed mixing between the singlet and doublet and the singlet
fermion $\psi_s\simeq \chi_1$ is the lightest of the neutral fermions.

In the limit
\begin{equation}
  |y_u|,\,|y_d| ~\ll~ 1, \qquad |m_s| ~\ll~ |\mu|,
\label{eq:lim}  
\end{equation}
we can expand the mass eigenvalues at the first order in  $y_{u,d}^2$ and get\,\footnote{An approximate expression of the corresponding rotation matrix $U$ is given in Appendix \ref{app:masses}.}
\begin{eqnarray}
\label{eq:mlim}  
&&m_{\chi_1} = m_s +\frac{v^2}{4} \frac{(y_u-y_d)^2}{\mu+m_s}-\frac{v^2}{4} \frac{(y_u+y_d)^2}{\mu-m_s} ,\\
&&m_{\chi_2} = -\mu -\frac{v^2}{4} \frac{(y_u-y_d)^2}{\mu+m_s}, \nonumber \\
&&m_{\chi_3} = \mu +\frac{v^2}{4} \frac{(y_u+y_d)^2}{\mu-m_s}. \nonumber
\end{eqnarray}
From these expressions, we see that, in the feebly-coupled regime, there is 
one neutral state of mass approximately $m_s$ (corresponding to $\chi_1 \sim \psi_s$), two
neutral fermions $\chi_{2,3}$ with mass approximately $\mu$ (equal
mixture of $\psi_u^0$ and $\psi_d^0$), and one charged fermion with
mass $m_\psi=\mu$, that we denote with $\psi^{\pm}$.
\begin{figure}[t]
\centering
\includegraphics[width=0.4\textwidth]{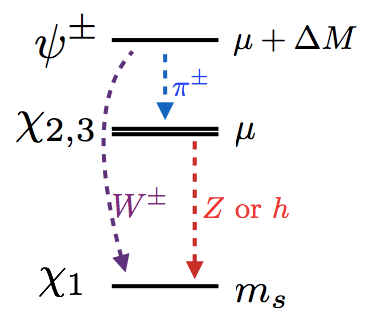}
\caption{\label{fig:spectrum} Mass spectrum of the Singlet-Doublet
  model considered here and possible decay modes.}
\end{figure}

The set of states with tree-level mass $\mu$ is further split by quantum corrections at one loop, which increases the mass of the charged state. 
Using the results of \cite{Cirelli:2009uv}, one finds that the splitting between the charged and the neutral states is
\begin{align}
&\Delta M=|m_{\psi}| -| m_{\chi_{2,3}}|=\frac{\alpha_2 \mu}{4 \pi} \sin^2 \theta_W f \left(\frac{m_Z}{\mu}\right), \\
& f(x)=\frac{x}{2}\left(2 x^3 \log x -2 x +\sqrt{x^2-4}(x^2+2)\log\left[\frac{1}{2}(x^2-2-x\sqrt{x^2-4})\right]\right),
\nonumber
\end{align}
where $\alpha_2 = g^2/(4\pi)$ (with $g$ being the $SU(2)_L$ gauge coupling) and $\theta_W$ is the weak mixing angle.
Considering $|\mu| >100$ GeV, $\Delta M$ spans the following range 
\begin{equation}
\label{eq:range}
250\,{\rm MeV} \lesssim \Delta M \lesssim 350\,{\rm MeV}\,.
\end{equation}
A sketch of the spectrum of the model and the possible decay
modes, described in the next section, is shown in Figure
\ref{fig:spectrum}.

\subsection{Decay modes and decay lengths}
\label{sec:decays}
In this subsection we study the decay modes of the fermion mass
eigenstates in the feeble coupling regime that control the phenomenology of the model.
General expressions for the decay widths through the model's Yukawa interactions can be found
in the Appendix of Ref.~\cite{Calibbi:2015nha}, while here we specialise to the regime of (\ref{eq:lim}) making
use of the expression for the mixing matrix reported in Appendix \ref{app:masses}.

Throughout this work, we consider $|\mu|> m_W$ as a doublet mass lower than about 90 GeV
is excluded by searches for charged fermions performed at LEP (see \cite{Egana-Ugrinovic:2018roi}, for a recent reassessment).
In this regime, the heavy charged states $\psi^{\pm}$ can decay directly into the lightest mass eigenstate,
$\psi^\pm\to W^{\pm} \chi_1$, via the suppressed Singlet-Doublet
mixing, or to the heavier neutral states ($\sim$ neutral components of
the doublet) and a soft pion, $\psi^\pm\to\pi^{\pm} \chi_{2,3}$, via
gauge interactions. The latter decay mode occurs via an off-shell $W$
and, for the range of charged-neutral state mass splitting reported in
Eq.~(\ref{eq:range}), it  dominates over possible leptonic modes
involving $\ell^{\pm}\nu$ instead of $\pi^{\pm}$. Half of the decays into pions plus neutral
  states go to $\pi^\pm\chi_2$ and the other half to $\pi^\pm\chi_3$,
with the partial decay widths given by \cite{Thomas:1998wy}:
\begin{equation}
\label{eq:psi2pi}
\Gamma[\psi^{\pm} \to \pi^{\pm} \chi_{2,3}] = \frac{G_F^2}{2\pi} \cos^2 \theta_c ~\, f_{\pi}^2 \Delta M^3 \sqrt{1-\left(\frac{m_{\pi}}{\Delta M}\right)^2},
\end{equation}
where $f_{\pi} \simeq 130$ MeV, $G_F$ is the Fermi constant and $ \theta_c$ the Cabibbo angle.  
As mentioned above, this decay mode
competes with the decay into the lightest fermion eigenstate $W^{\pm}
\chi_1$ induced by the small Singlet-Doublet mixing.  At leading order
in the Yukawa couplings, taking $\tan\theta=1$
(i.e.~$y_u=y_d=\frac{y}{\sqrt{2}}$), the decay width reads
(for $\mu \gtrless 0$):
\begin{align}
\Gamma[\psi^{\pm} \to W^{\pm}\chi_1] =& \frac{\alpha\, y^2 v^2}{32 s_{W}^2} ~\sqrt{\lambda(m_{\psi}^2,m_{\chi_1}^2,m_W^2)}~\times \nonumber\\
&\frac{\left((m_{\psi}\pm m_{\chi_1})^2+2 m_W^2\right)\left((m_{\psi}\mp m_{\chi_1})^2-m^2_W\right)}{m_{\psi}^3 m_W^2 (\mu-m_s)^2} ,
\label{eq:psi2W}
\end{align}
 where
 $$\lambda(a,b,c)= a^2+b^2+c^2-2 ab -2 ac -2 bc,$$  
$\alpha$ is the electromagnetic constant, and $s_W\equiv \sin\theta_W$. 
It can be shown that $\Gamma[\psi^{\pm} \to W^{\pm}\chi_1]$ has a negligible dependence on $\tan\theta$ in the
limit~(\ref{eq:lim}), and hence the formula \eqref{eq:psi2W} will suffice to our purposes.
\begin{figure}[t]
  \centering
 \includegraphics[width=0.45\textwidth]{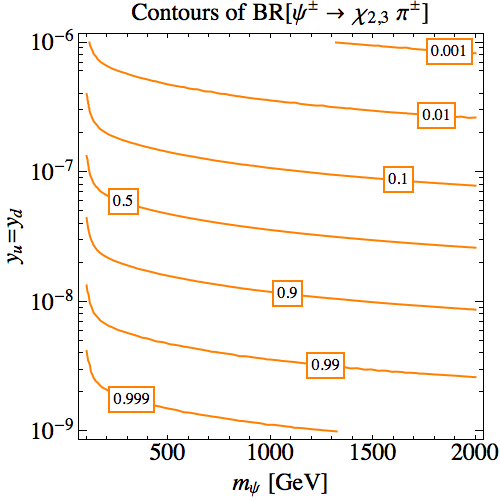}\hfill
 \includegraphics[width=0.45\textwidth]{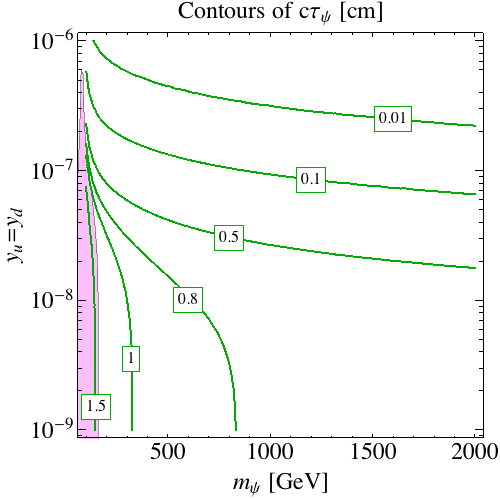}
\caption{\label{fig:BRchargino} Left: branching ratios of the decay of
  the charged fermion into $\chi_{2,3}$ plus a pion. Right: decay
  length (in cm) of the charged fermion; the shaded region is excluded by the
  ATLAS search for disappearing tracks \cite{Atlas-DT,Atlas-DT2}.  The
  mass of the lightest neutral fermion is fixed to $m_{\chi_1} =10$
  keV. 
  }
\end{figure}

Making use of Eqs.~(\ref{eq:psi2pi}) and~(\ref{eq:psi2W}), we show in
Figure~\ref{fig:BRchargino} the branching ratios for the decay process
$\psi^{\pm} \to \pi^{\pm} \chi_{2,3}$ (left panel) and the contours
for fixed values of the charged fermion decay length $c \tau_\psi$
(right panel) on the $(m_\psi,\,y)$ plane in the custodial symmetry
limit $y_u=y_d$ for $m_{\chi_1}= 10$ keV.  We can see that for $y \leq
10^{-8}$ the decay mode into $\pi^{\pm} \chi_{2,3}$ is the dominant
one. Moreover, comparing the two panels of
Figure~\ref{fig:BRchargino}, we can see that, when $\psi^\pm$ decays
preferably into pions plus heavy neutral state, the decay length is
about 1 cm that is approximately the minimal length, to which LHC searches for disappearing charged tracks have sensitivity.
For illustrative purposes, we
thus also show, in the right panel of Figure~\ref{fig:BRchargino}, a
purple region excluded by a recent ATLAS analysis
\cite{Atlas-DT,Atlas-DT2}.  We will address the collider constraints
in more details in Section~\ref{sec:LHC}.  Let us mention that the
overall picture depends neither on $m_{\chi_1}$ nor on $\tan \theta$
in the limit~(\ref{eq:lim}) that we are interested in here.

\begin{figure}[t]
  \centering
 \includegraphics[width=0.45\textwidth]{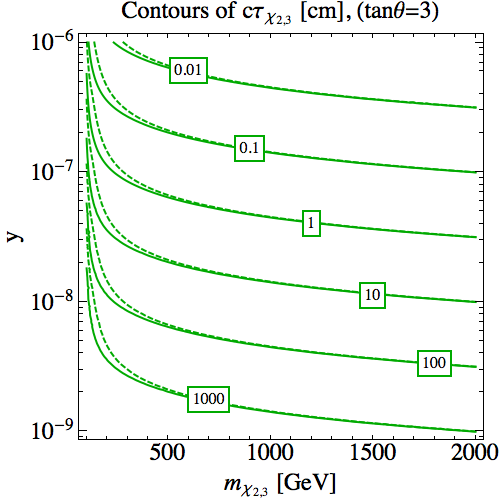}\hfill
  \includegraphics[width=0.45\textwidth]{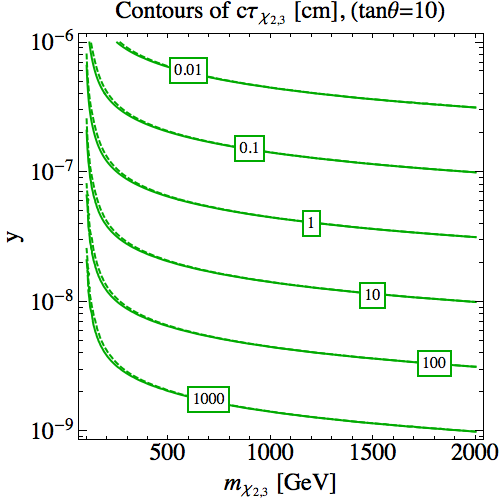}
\caption{\label{fig:decaylength} Decay length of the two heavy neutral
  fermions, $\chi_2$ (solid lines) and $\chi_3$ (dashed lines) for two
  different choices of tan $\theta= y_u/y_d$.  The
  mass of the lightest neutral fermion is set to $m_{\chi_1} =10$
  keV. }
\end{figure}

The two neutral states $\chi_2$ and $\chi_3$ can decay either into $Z
\chi_1$ or into $h \chi_1$ through the Yukawa interactions.  At leading order
in $y_u$ and $y_d$, the decay widths read (for $\mu \gtrless 0$): \bea
\label{decay_neutr}
\Gamma[\chi_2 \to Z \chi_1 ] &=& \frac{\alpha\, (y_u+y_d)^2}{64 s_W^2 c_W^2} ~\sqrt{\lambda(m_{\chi_2}^2,m_{\chi_1}^2,m_Z^2)}~\times\nonumber \\
&&\frac{v^2 \left((m_{\chi_2}\pm m_{\chi_1})^2+2m_Z^2\right)\left((m_{\chi_2}\mp m_{\chi_1})^2-m_Z^2\right)}{m_{\chi_2}^3 m_Z^2 (\mu-m_s)^2} , 
 \\
\Gamma[\chi_3 \to Z \chi_1 ]& =& \frac{\alpha\, (y_u-y_d)^2}{64 s_W^2 c_W^2} ~\sqrt{\lambda(m_{\chi_3}^2,m_{\chi_1}^2,m_Z^2)} ~\times \nonumber \\
&&\frac{v^2 \left((m_{\chi_3}\mp m_{\chi_1})^2+2m_Z^2\right)\left((m_{\chi_3}\pm m_{\chi_1})^2-m_Z^2\right)}{m_{\chi_3}^3 m_Z^2 (\mu+m_s)^2} ,
\\
\Gamma[\chi_2 \to h \chi_1]& =& \frac{(y_u-y_d)^2}{64 \pi} \frac{(m_{\chi_2}\mp m_{\chi_1})^2 -m_h^2}{m_{\chi_2}^3} ~ \sqrt{\lambda(m_{\chi_2}^2, m_{\chi_1}^2, m_h^2)},  \\
\Gamma[\chi_3 \to h \chi_1] &=& \frac{(y_u+y_d)^2}{64 \pi} \frac{(m_{\chi_3}\pm m_{\chi_1})^2 -m_h^2}{m_{\chi_3}^3}  ~\sqrt{\lambda(m_{\chi_3}^2, m_{\chi_1}^2, m_h^2)} , 
\label{decay_neutr4}
\eea
after having redefined the fermionic fields in order for $m_{\chi_i}$
to be positive. Notice that, in practice, in the regime
of~(\ref{eq:lim}), the above decay widths are not affected by the sign
of $\mu$.  The typical decay lengths of $\chi_2$ (solid line) and
$\chi_3$ (dashed line) are shown as a function of their mass in
Figure~\ref{fig:decaylength} for two values of $ \tan\theta$ and
$m_{s} = 10$~keV.  As can be seen, for a $y$ coupling of the order
$10^{-7}$, the decay length is around 1 cm, while it exceeds $10$ m
for $y \sim 10^{-9}$.  In addition, the decay lengths of $\chi_2$ and
$\chi_3$ appear to be essentially equal for $m_{\chi_{1,2}} \gtrsim
300$ GeV or for $\tan\theta\gg 1$.  Let us also emphasize that, in the
limit of Eq.~(\ref{eq:lim}), the decays $\chi_3\to\chi_2$ are not
allowed due to the tiny mass splittings, as it can be verified by
inspecting the expressions in Eq.~(\ref{eq:mlim}).

\section{Dark matter abundance from Freeze-in }
\label{sec:FI}
In the feeble Yukawa coupling regime that we are considering, the dark
matter candidate $\chi_1\sim \psi_s$
has strongly suppressed interactions with the SM particles. Hence it is 
not in thermal equilibrium with the SM bath at the time of
production. In contrast, the components of the electroweak doublet are
in thermal equilibrium because of their unsuppressed gauge
interactions.  Assuming zero initial abundance of $\chi_1$, the dominant
production mechanism for the dark matter particle is through the decay
of the heavy mediators ($\psi^{\pm}$ and $\chi_{2,3}$) along the
cosmological evolution.\footnote{In our model, the scattering processes with $\chi_1$ in the final state give definitely subdominant contributions with respect to mediators' decays, such as in e.g.~\cite{Hall:2009bx}.
Moreover, decays of SM particles into DM, such as $Z\to\chi_1 \chi_1$ and $h\to \chi_1 \chi_1$, also give negligible contributions, because doubly suppressed by the tiny Singlet-Doublet mixing.} 
This production ``freezes-in'' when the
abundance of the heavy mediators is Boltzmann suppressed, that is
approximately when the temperature drops below their mass. This is the
framework in which we carry out our analysis; see
e.g.~\cite{Hall:2009bx,Frigerio:2011in,Chu:2011be,Bernal:2017kxu,Frere:2006hp,Co:2015pka,Hessler:2016kwm,Brooijmans:2018xbu}
for some previous examples. The DM comoving number density
induced through the decay of $A \to B\, \chi_{\rm DM}$ simply
reduces to \cite{Hall:2009bx}
\begin{equation}
\label{FI_Hall}
Y_{\chi_{\rm DM}} = \frac{135 g_A}{ (1.66) 8 \pi^3 g_*^{3/2}} \frac{M_{Pl} \Gamma_A}{m_A^2}\,,
\end{equation}
where $g_A$ counts the spin degrees of freedom of the mother particle
$A$, $g_*$ is the number of degrees of freedom at the freeze-in
temperature $T \sim m_A$, and $M_{Pl}= 1.22 \times 10^{19}$ GeV is the
Planck mass. This result is obtained making the following simplifying
assumptions: (i) the mother particle $A$ and the daughter particle $B$
are in thermal equilibrium with the SM thermal bath; (ii) $A$ follows
a Maxwell-Boltzmann distribution function; (iii) we can neglect the
Pauli-blocking/stimulated emission effects associated to $B$. See also
the discussion in~\cite{Belanger:2018ccd}.

Let us notice that the freeze-in mechanism considered here\,---\,taking place through the decay
of a mediator that is in thermal equilibrium with the Standard Model sector
in the early universe\,---\,shares some similarities with the so-called superWIMP
mechanism \cite{Feng:2003xh}. In the latter case, the late decay of the WIMP mother particle\,---\,occurring considerably after its freeze-out\,---\,give
rise to the dark matter abundance. The main difference between the two mechanisms is the life-time of the
mother particle in thermal contact with the Standard Model. In the
superWIMP case, the life-time is typically much longer ($\tau\sim 10^5-10^8$ s) and the scenario
is subject to constraints from big bang nucleosynthesis~\cite{Feng:2003xh,Feng:2003uy}.

\subsection{Freeze-in: collider and cosmology interplay }
\label{sec:coll-cosm-interpl}

In the context of the Singlet-Doublet model, three mass degenerate
heavy states can decay into dark matter, namely the charged fermion
$\psi^{\pm}$ and the two neutral fermions $\chi_{2,3}$, giving rise
to the DM yield:
\begin{equation} 
\label{eq:yield}
Y_{\chi_1}= \frac{270 M_{Pl}}{ (1.66) 8 \pi^3 g_*^{3/2}}  \left( \sum_{B=Z,h} \frac{\Gamma[\chi_3 \to B \chi_1]}{m_{\chi_3}^2} + 
\sum_{B=Z,h} \frac{\Gamma[\chi_2 \to B \chi_1]}{m_{\chi_2}^2}  + g_\psi \frac{\Gamma[\psi^{+} \to W^+ \chi_1]}{m_{\psi}^2}\right),
\end{equation}
where $g_{\psi}= 2$ takes into account the number of degrees of freedom of the charged
fermion.
Notice that the contributions of the heavy neutral fermions $\chi_{2,3}$
directly depend on the total decay widths of $\chi_{2,3}$, which
we will denote $\Gamma_{\chi_{2,3}}$ in the following, as the decays into $\chi_1$ are
the only available decay modes.  On the other hand, for the
  charged fermion, only the partial decay width into the $W^+ \chi_1$
  final state appears since the $\psi^\pm$ decays into $\chi_{2,3}$ are
  already accounted for by the two first contributions associated to
  the thermal equilibrium abundances of $\chi_{2,3}$.   
We can now compute the DM relic density in terms of $Y_{\chi_1}$:
\begin{eqnarray}
\label{abundance_gen}
\Omega_{\chi_1} h^2=m_{\chi_1}\, \frac{s_0 h^2}{\rho_c} Y_{\chi_1}, 
\end{eqnarray}
where the present entropy density and critical density are respectively 
$s_0=2.8912 \times10^9$ m$^{-3}$ and $\rho_c=10.537 \, h^2$~GeV/m$^3$. 
Considering that, in our scenario, we have $m_{\chi_{2,3}}\simeq m_\psi \simeq \mu$, 
we obtain as a result
\begin{equation}
\label{abundance_gen2}
\Omega_{\chi_1} h^2  = 0.1 \left(\frac{105}{g_*} \right)^{3/2} \left(\frac{m_{\chi_1}}{10 \text{~keV}} \right) \left( \frac{1 \text{~TeV}}{\mu} \right)^2
\left( \frac{ \sum_{ij}g_{A_i}\Gamma_{ij}}{5\times10^{-15} \text{~GeV}}\right), 
\end{equation}
where $g_{A_i}$ is the number of degrees of freedom of the mother
particle $A_i$, $\Gamma_{ij}$ denotes the decay width $\Gamma[A_i\to B_j
  \chi_1]$, with $ A_i=\chi_{2,3}^0$ or $\psi^{\pm}$ decaying into
$\chi_1$ plus a SM boson, $B_j= Z,\,h$ or $W^\pm$.  

Obtaining the dark matter yield on more general grounds, starting
from Maxwell-Boltzmann statistics, requires a fully numerical
treatment of the evolution equations, which makes the computation and the
interpretation of the freeze-in mechanism less straightforward. The
authors of Ref.~\cite{Belanger:2018ccd} have however recently
delivered the public code ${\tt micrOMEGAs5.0}$ that allows to easily
handle such computations. We have explicitly checked that the analytical results presented here
are in excellent agreement with the ones obtained with ${\tt micrOMEGAs5.0}$ (employing the Singlet-Doublet model files from our implementation in {\tt
  FeynRules}~\cite{Alloul:2013bka}) in the Maxwell-Boltzmann limit.\footnote{
  See Ref.~\cite{Belanger:2018ccd} for a discussion on the relevance of
  the statistics in different dark matter scenarios.} 
  Beyond this simplifying assumption, the full numerical treatment of the evolution
equations gives rise to a moderate positive correction to $\Omega_{\chi_1} h^2$ with respect to
our analytical result (about 25\%).
In this paper, we choose to discuss the results of our analysis with the
Maxwell-Boltzmann approximation, thus neglecting the above small correction, 
in order to have a fully analytical understanding of the parameter space of the model
yielding the observed DM abundance.

The result of Eq.~\eqref{abundance_gen2} is rather generic for the
freeze-in scenarios, independently of the underlying
dark matter model, and indicates the typical order of magnitude of the physical quantities involved, that is
the dark matter mass, the mediator(s) mass, and the mediator(s) widths.  
Note that the decay length of a
particle is related to the total decay width through
\begin{equation}
c \tau_{A} = \frac{10^{-15} \text{~GeV}}{\Gamma_{A}} \times 19 \text{~cm} \,. 
\end{equation}
Hence mother particles or equivalently mediators, $A$, with
a total decay width allowing for the dark matter density to be in
agreement with the observed abundance $\Omega h^2\simeq 0.12$, are in
the right ballpark to give rise to macroscopically long displacements
at colliders. In order for the mediators to be produced at the LHC,
their mass cannot typically exceed the TeV scale.  For such mass
scale, a DM mass in the heavy range $1\,\text{MeV} \lesssim m_{\chi_1} \lesssim
1\,\text{GeV}$ corresponds to mediators escaping the detectors, while for
light dark matter, $1\,\text{keV} \lesssim m_{\chi_1} \lesssim 1\,\text{MeV}$, the
signature will be characterised by displaced vertices visible inside
the detectors. 
This highlights the natural interplay among LHC
long-lived or displaced signatures, the freeze-in mechanism, and
cosmological or astrophysical probes of light ($\simeq$ keV) dark
matter.
An important remark is that these considerations and
correlations are strictly correct for mediators that can  decay
into dark matter only, which is the case of $\chi_{2,3}$ here.  We study
this complementarity in further detail in Sections \ref{sec:LHC} and
\ref{sec:discussion}.

\subsection{The viable dark matter parameter space }
\label{sec:coll-cosm-interpl}
\begin{figure}[t]
  \centering
 \includegraphics[width=0.45\textwidth]{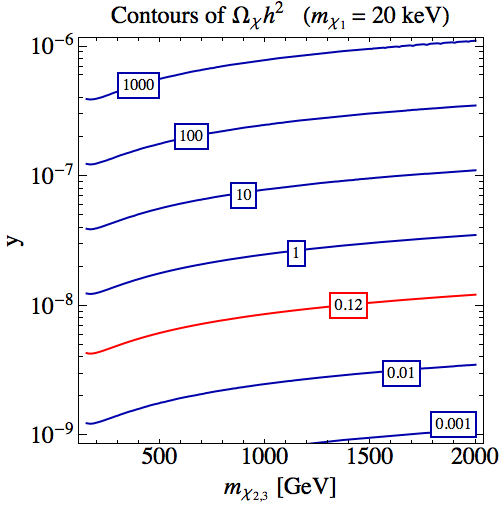}\hfill
  \includegraphics[width=0.45\textwidth]{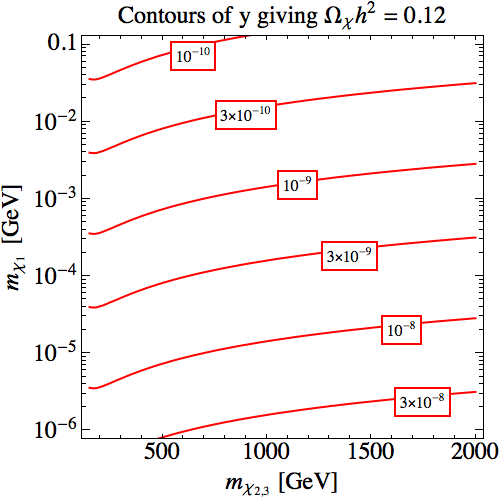}
\caption{\label{fig:Omega}
Left: Contours of $\Omega h^2$ for $m_{\chi_1}=m_s = 20$ keV. Right: Contours of the values of $y$ on the $(m_{\chi_{2,3}}\,,m_{\chi_{1}})$ plane required to get the observed DM relic abundance.
}
\end{figure}

We can now employ the decay widths that we computed in
Section~\ref{sec:decays} to derive the predicted value of the dark
matter relic abundance on the parameter space of the Singlet-Doublet
model.  Expanding the expressions of the decay widths
in the limit of small $m_{\chi_1}$ one finds at leading order:
\bea
&&
\sum_{i=2,3}\Gamma[\chi_{i} \to Z \chi_1] = 
\frac{y^2}{32 \pi} \frac{ \left(m_Z^2-\mu^2 \right)^2  \left( 2 m_Z^2+\mu^2 \right) }{ \mu^5}\\
&&
\sum_{i=2,3}\Gamma[\chi_{i} \to h \chi_1] = 
\frac{y^2}{32 \pi} \frac{\left(m_h^2-\mu^2 \right)^2  }{\mu^3}\\
&&
\Gamma[\psi^{\pm} \to W^{\pm} \chi_1] = 
\frac{ y^2}{32 \pi} \frac{ \left(m_W^2-\mu^2 \right)^2  \left( 2 m_W^2+\mu^2 \right) }{ \mu^5}\, .
\eea
These expressions show that the combinations of the decay widths entering in the computation of the relic
abundance do not depend on $\tan \theta$ at zeroth order in $m_{\chi_1}$.  
Plugging these expressions into
Eq.~\eqref{abundance_gen},  we find the following approximate expression
\begin{equation}
  \Omega_{\chi_1} h^2 \simeq 0.11 \left( \frac{105}{g_*} \right)^{3/2} \left(\frac{y}{10^{-8}} \right)^2   \left(\frac{m_{\chi_1}}{10 \text{~keV}} \right) \left(\frac{700 \text{~GeV}}{\mu} \right),
  \label{eq:simpleOm}
\end{equation}
which accounts for the correct relic abundance up to a few percent
level error when $\mu \gtrsim 400$ GeV.  Eq.~(\ref{eq:simpleOm})
shows how the dark matter relic
abundance via freeze-in scales with the different parameters of the
model. The results of the dark matter calculations
presented in what follows always make use of the full expressions of
Eqs.~(\ref{eq:yield},\,\ref{abundance_gen}) with $g_* =105$.

In the left panel of Figure~\ref{fig:Omega}, we show the dependence of
the dark matter abundance through the freeze-in mechanism on the
parameters of the model for a fixed DM mass $m_{\chi_1} = 20$ keV on
the ($m_{\chi_{2,3}}$, $y$) plane, or equivalently ($\mu$, $y$) plane.
It appears that, for $m_{\chi_1}\sim$ few tens of keV, the coupling
should be $y = {\cal O} (10^{-8})$ in order to reproduce the observed
dark matter density.  In the right panel of~Figure~\ref{fig:Omega}
instead, we show on the ($m_{\chi_{2,3}}$, $m_{\chi_{1}}$) plane
contours of the values of the coupling $y$ that yield $\Omega_{\chi_1}
h^2 =0.12$.  We can see that, for $m_{\chi_1}$ in the [100 MeV, 1 keV]
mass range and a $\mu$ scale of relevance for colliders, the required
size of the Yukawa coupling is in the range
\begin{equation}
\label{window_y}
10^{-8} \lesssim y \lesssim 10^{-10}\,.  
\end{equation}
The largest values of the coupling $y\sim 10^{-8}$ allow for very
light dark matter candidates (few keV) to account for all the dark
matter while heavier particles of hundreds of MeV requires even more
suppressed Yukawa interactions with $y\sim 10^{-10}$. This observation
will be relevant when comparing the reach of collider experiments to
the one of cosmology probes on the frozen-in Singlet-Doublet dark
matter parameter space.

\section{Cosmology probe of light dark matter}
\label{sec:cosmo-bounds}
 Dark matter candidates with non negligible velocity dispersion deep
 in the radiation dominated era can leave a distinctive imprint in
 cosmology and astrophysics observations due to their free-streaming
 that delays the structure growth.  Overdensities are suppressed below the
 comoving free-streaming horizon given by
\begin{equation}
  \lambda_{fs}=\int_{0}^1 \frac{\langle v\rangle}{a^2 H}da
\label{eq:lfs}
\end{equation}
where $a$ is the scale factor and $H$ is the Hubble rate and $\langle
v\rangle$ is the velocity dispersion of the dark matter ($\langle
v\rangle$ is given by the velocity of light for relativistic dark matter). 
For example, thermal warm dark matter (WDM), which was
in thermal equilibrium and relativistic until decoupling at
temperature $T_D$, has a free-steaming length of $\lambda_{fs}\simeq$
Mpc $($keV$/m_X) \, T_D/T_\nu$ where $T_\nu$ is the temperature of active
neutrinos and $m_X$ is the WDM mass.\footnote{ The thermal abundance
  of WDM is given by $\Omega_X h^2\simeq (T_D/T_\nu)^3 m_X/94$ eV
  where $T_D/T_\nu= \left(g_*(T_\nu)/g_*(T_D)\right)^{1/3}$ for
  entropy conservation with $g_*(T)$ the effective number of
  relativistic degrees of freedom and $g_*(T_\nu) = 10.75$. All
  together, an injection of large number of relativistic degrees
  ($>10^4$) of freedom is needed, compared to the available ones in
  the SM $g_*(T>T_{EW})=106.75$, so as to be able to get a thermal WDM
  of a still allowed few keV
  WDM~\cite{Irsic:2017ixq,Yeche:2017upn,Viel:2013apy}.} Such a WDM
scenario has served as a benchmark for non-cold dark matter
cosmology. In this work we exploit the results of Lyman-$\alpha$
forest studies that have been used to set a lower bound on the thermal
WDM mass of
\begin{equation}
  m_{\rm WDM}>4.65 \mbox{ keV \qquad [ thermal WDM ]}\,
  \label{eq:mwdmly}
\end{equation}
at 95\% confidence level (CL)~\cite{Yeche:2017upn}.  Notice that the
above constraint can e.g.~relax to $m_{\rm WDM}>3.2$ keV at 95\% CL when
the WDM makes only part ($>80\%$) of the total DM
content~\cite{Baur:2017stq}. It has also been argued that considering
a different thermal history in the treatment of the Lyman-$\alpha$
forest data (especially the ones associated to high redshift quasars),
a few keV DM candidate could even provide a good fit to the data,
see~\cite{Viel:2013apy, Garzilli:2015iwa,Baur:2017stq}, a possibility
that is strongly challenged by X-ray constraints in the context of
sterile neutrinos~\cite{Baur:2017stq}.

Thermal warm dark matter is not the only relic that would suppress
small scale structure formation. Other DM candidates with non
negligible velocity at the time of production will give rise to
similar effect.\footnote{Notice that collisional or Silk damping (in
  contrast with free streaming $\equiv$ collision-less damping) can
  also give rise to a similar imprint in small scale structures. This
  would be typically the case of dark matter interacting with
  relativistic species, see
  e.g.~\cite{Moline:2016fdo,Schewtschenko:2015rno,Schewtschenko:2014fca,Wilkinson:2014ksa,Murgia:2017lwo,Buckley:2014hja,Vogelsberger:2015gpr,Cyr-Racine:2015ihg,Cyr-Racine:2013fsa,Lovell:2017eec,Escudero:2018thh}.}
Among them, one finds (non-)resonantly produced sterile
neutrinos~\cite{Dodelson:1993je, Shi:1998km,Dolgov:2000ew,
  PhysRevD.64.023501,1126-6708-2007-01-091}, sterile neutrinos from
frozen-in scalars~\cite{Merle:2013wta,Kang:2014cia}, mixed dark matter
scenarios~\cite{Boyarsky:2008xj} and\,---\,of interest for this paper\,---\, other frozen-in
particles~\cite{Hall:2009bx,Heeck:2017xbu,Bae:2017dpt,Boulebnane:2017fxw,Brdar:2017wgy}. The
different mechanisms of production involved can typically give rise to
distribution functions that can differ from the (thermal) Fermi-Dirac
distribution. As a result, the imprint on the linear matter power
spectrum should a priori be recomputed making use of the relevant
Boltzmann codes. Dedicated hydrodynamical simulations should then
be performed so as to extract the non-linear evolution of a baryon+DM
population and properly compute the observables relevant to estimate
the Lyman-$\alpha$ flux power spectra within a given DM scenario and
compare with data. All this procedure is however beyond the scope of
this paper.

\begin{figure}[t]
  \centering
 \includegraphics[width=0.5\textwidth]{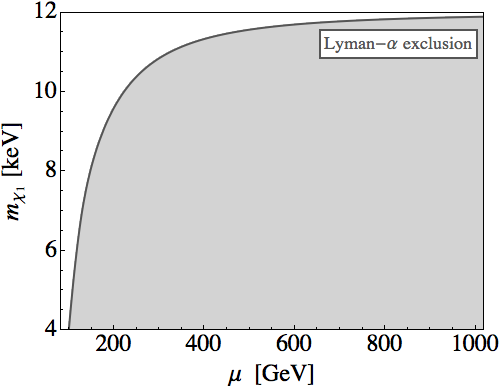}
\caption{\label{fig:lybound}
Bound on the DM mass from Lyman-$\alpha$; the area below the curve is excluded.}
\end{figure}

Here we use the constraints that have been derived
in~Refs.~\cite{Heeck:2017xbu,Boulebnane:2017fxw} on keV dark matter
produced through the freeze-in from the decay(s) of some thermalised
mother particle $A$ into the DM and another daughter particle $B$.
In~Ref.~\cite{Boulebnane:2017fxw}, the suppression of the linear
matter power spectrum in the freeze-in scenario has been computed and
compared to the one of thermal WDM with a mass of 4.65
keV.\footnote{In Ref.~\cite{Boulebnane:2017fxw}, the transfer function of
  DM (associated to the ratio of cold DM and freeze-in DM linear matter
  power spectra) produced through freeze-in from the decays of some
  thermalised mother particle $A\to B +$ DM always appear to have the very
  same spectral form as the one of thermal WDM. For other references,
  estimating the range of viable non-cold dark matter candidates based
  on the derivation of the linear matter power spectrum, see
  e.g.~\cite{Boyarsky:2008xj,Merle:2015vzu,Merle:2015oja,Konig:2016dzg,Bae:2017dpt,Murgia:2017lwo}
} This provided a constraint on the mass of the frozen-in dark matter
particle that shows a dependence on the mass splitting between $A$ and
$B$ for compressed spectra. Within our framework, the resulting
constraint on the dark matter mass reads:
\begin{equation}
  m_{\rm DM}> 12~{\rm keV} \left( \frac{ \sum_{ij}g_{A_i}\Gamma_{ij} \Delta_{ij}^\eta}{\sum_{ij}g_{A_i} \Gamma_{ij}}\right)^{1/\eta}\,.
  \label{eq:mlyheeck}
\end{equation}
where $\eta=1.9$ (as obtained in Ref.~\cite{Boulebnane:2017fxw} from a
numerical fit), and $\Gamma_{ij}$ is the decay width $\Gamma(A_i\to
B_j \chi_1)$, $\Delta_{ij}= (m_{A_i}^2-m_{B_j}^2)/m_{A_i}^2$ with $
A_i=\chi_{2,3}^0$ and $\psi^{\pm}$ denoting the mediators that decay
into the dark matter fermion $\chi_1$ and another SM final state $B_j=
Z,\,h$ or $W^\pm$.  The results of Ref.~\cite{Boulebnane:2017fxw} imply
thus that, in general, frozen-in DM, resulting from the decay of a
thermalised mother particle with $m_{\rm DM}>$ 12 keV, evades the
constraints from the Lyman-$\alpha$ forest data derived
in~\cite{Yeche:2017upn}. Lower DM masses can become allowed when
$\Delta_{ij}$ is small, {\it i.e.} for small mass splittings between
$A_i$ and $B_j$ (as pointed out in \cite{Heeck:2017xbu}).  Our bound is
shown in Figure~\ref{fig:lybound} as a function of the doublet mass
scale $\mu$. As we can see, the lower bound on the DM mass becomes
weaker than $m_{\chi_1}>12$ keV only for values of the doublet mass $\mu$ approaching the mass of the decay products $B_j=
Z,\,h,\,W^\pm$.

Notice that astrophysics and cosmology already provide other
complementary probes of dark matter scenarios suppressing structure
formation on small scales. Among them one finds CMB fluctuations, galaxy clustering, galaxy satellite number
count, etc., see
e.g.~Refs.~\cite{Viel:2005qj,Seljak:2006qw,Viel:2006kd,Murgia:2017lwo,2018MNRAS.473.2060J,Vogelsberger:2015gpr,
  Lopez-Honorez:2017csg,Lovell:2017eec,Escudero:2018thh,
  Villanueva-Domingo:2017lae}; and also
e.g.~Refs.~\cite{Sitwell:2013fpa,Carucci:2015bra,Carucci:2016yzq,Lovell:2017eec,Escudero:2018thh}
for future probes.  Currently, most (combinations of) probes tend to
exclude a few keV thermal warm dark matter scenarios, on a par with
the results of the Lyman-$\alpha$ forest analysis considered here.
 
\section{Signatures at the LHC}
\label{sec:LHC}
\begin{figure}[t]
  \centering
 \includegraphics[width=0.6\textwidth]{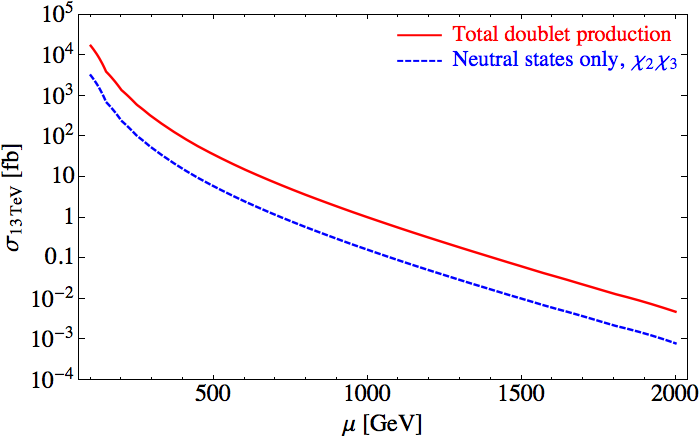}
\caption{\label{fig:xsec}
NLO production cross section of the states belonging to the fermion doublet pair at the LHC with $\sqrt{s}$=13 TeV as computed by {\tt Prospino2} \cite{Beenakker:1999xh}. The solid red line indicates the sum over all possible production modes, while the dashed blue line shows the production cross section of the neutral states $\chi_2\chi_3$ only.}
\end{figure}

In the feebly-coupled regime of the Singlet-Doublet dark matter model,
the mediators $\psi^{\pm}$ and $\chi_{2,3}$ are essentially the
charged and neutral components of the extra $SU(2)_L$ doublets. As a
result, they can be produced at the LHC through electroweak 
processes. These production processes are induced by gauge couplings only and
thus the cross sections are independent of the couplings
$y_u$ and $y_d$. They are actually equal to those of a pure Higgsino
in supersymmetry (SUSY) that can be computed using public
  tools such as {\tt Prospino2}~\cite{Beenakker:1999xh}.  Referring
to the SUSY nomenclature, the relevant production modes include
neutralino pair production, chargino pair production, and associated
production of neutralino and chargino:
\begin{equation}
pp \to \chi_2 \chi_3 + X,\quad pp \to \psi^+ \psi^- + X,\quad pp \to \chi_{2,3} \psi^\pm +X. \label{eq:modes}
\end{equation}
Being substantially decoupled from the SM sector, 
the singlet dark matter $\chi_1$ can only be produced at the last step of
the decay chain, with the possible decay modes as illustrated in
Figure~\ref{fig:spectrum}. We report in
Figure~\ref{fig:xsec} the total production cross section (obtained by summing over
all mediator pair and associated production modes) with a 
continuous red line, and the production cross section of a $\chi_2 \chi_3$ pair with a dashed blue
 line, as a function of the doublet mass scale $\mu$. The cross sections were computed by means of  
 {\tt Prospino2} for $pp$ collisions with $\sqrt{s}$\,=\,13 TeV at next-to-leading order (NLO). 
 \begin{figure}[t]
  \centering
 \includegraphics[width=0.45\textwidth]{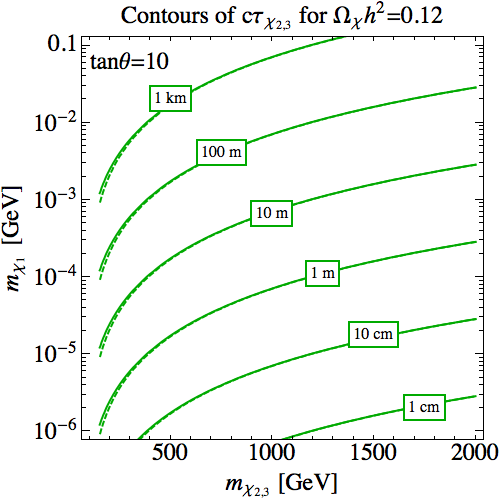}\hfill
  \includegraphics[width=0.45\textwidth]{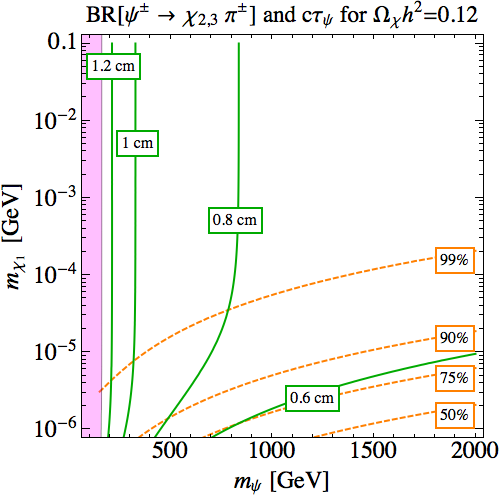}
\caption{\label{fig:chidecay} Left: Decay length of the neutral
  fermions $\chi_2$ (solid lines) and $\chi_3$ (dashed lines).  Right:
  Branching ratio (orange dashed lines) and decay length (green solid lines) of $\psi^{\pm}\to\chi_{2,3} \pi^{\pm}$;  the shaded region is excluded by searches for disappearing tracks \cite{Atlas-DT,Atlas-DT2}.  The coupling $y$ is set in both plots by requiring $\Omega h^2 =0.12$. }
\end{figure}

In order to obtain the collider constraints, we first have to compute
the typical decay length of the heavy mediators in the viable dark
matter parameter space, i.e.~where $\Omega_{\chi_1} h^2 = 0.12$. In
Figure~\ref{fig:chidecay}, we present the decay length of the
mediators for the model parameters accounting for the whole observed
dark matter abundance.  In the left panel, we show contours for the decay length
of the heavy neutral fermions, $c\tau_{\chi_{2,3}}$.  On general
grounds, the results depend on $\tan \theta$ but, as already noticed
in Figure~\ref{fig:decaylength}, the $\tan \theta$ dependence is
negligible as long as $\mu\gtrsim300$ GeV or $\tan\theta\gg1$.
Also, as expected from the discussion in
Section~\ref{sec:FI}, the figure shows that decay lengths
leading to displaced signatures within the volume of LHC detectors
correspond to the light dark matter regime, $m_{\chi_1}\lesssim 1$
MeV. On the right panel of Figure~\ref{fig:chidecay}, the dashed
orange contours indicate the branching fraction of the $\psi^\pm$ decay
into pions and $\chi_{2,3}$. We see that this decay mode is dominant
except in a small corner of the parameter space where $m_{\chi_1} = \ord{1}$
keV and $m_\psi=\mu$ is larger than about 1~TeV. The $\psi^\pm$ decay
mode into $W^\pm\chi_1$, driven by the Yukawa interactions and
contributing to the dark matter relic abundance, is typically
subdominant, due to the feeble couplings involved. As a consequence,
the decay length of $\psi^\pm$ is always of the order of 1 cm in the
parameter region relevant for the freeze-in mechanism as shown by the
green solid lines in the right panel of Figure~\ref{fig:chidecay}.
\begin{figure}[t]
  \centering
 \includegraphics[width=0.45\textwidth]{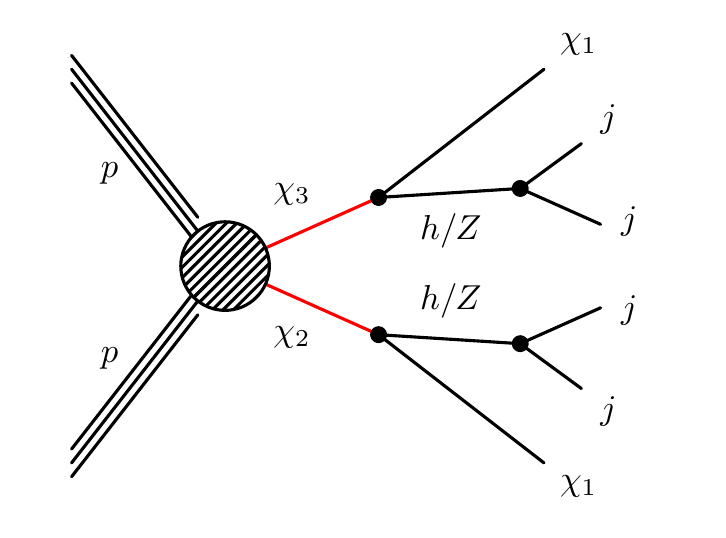}
\caption{\label{fig:feyn} Schematic representation of one of the
  processes leading to displaced $Z$ or $h$ bosons plus missing energy
  at the LHC. The red lines denote long-lived particles. Similar final
  states arise from $\chi_2$ or $\chi_3$ pair production. Note that we
  do not specify the production mechanism of the pair of neutral heavy
  fermions since it could be produced directly through electroweak
  processes or through the decay of the charged fermion. }
\end{figure}

We can now discuss the LHC signatures of our freeze-in Singlet-Doublet model.
\\[2mm] 
\noindent $\bullet$ {\bf Disappearing tracks:} Independently of the final
steps of the decay chain, the charged fermions $\psi^{\pm}$ decay with
a small displacement (of the order $1$ cm at most, cf.~Figure~\ref{fig:chidecay}) leading to `disappearing' charged tracks that can be
searched for at the LHC. In fact, a recent ATLAS analysis \cite{Atlas-DT} (see also the similar search~\cite{Sirunyan:2018ldc} from the CMS Collaboration),
reinterpreted in \cite{Atlas-DT2} in terms of pure Higgsino
production (which is exactly our case), excludes the regions shaded in purple in
the right panels of Figs.~\ref{fig:BRchargino} and \ref{fig:chidecay}. 
This search constrains the mass of the charged fermions to be larger than about 150 GeV in the
regime in which $\psi^\pm\to\pi^{\pm} \chi_{2,3}$ dominates. For future prospects of searches for disappearing tracks and possible
strategies to increase their sensitivity, see \cite{Mahbubani:2017gjh,Fukuda:2017jmk},
where Higgsino masses up to approximately $400-500$ GeV are foreseen to be accessible at the future high-luminosity run of the LHC (HL-LHC).
\\[2mm] 
\begin{figure}[t]
  \centering
 \includegraphics[width=0.91\textwidth]{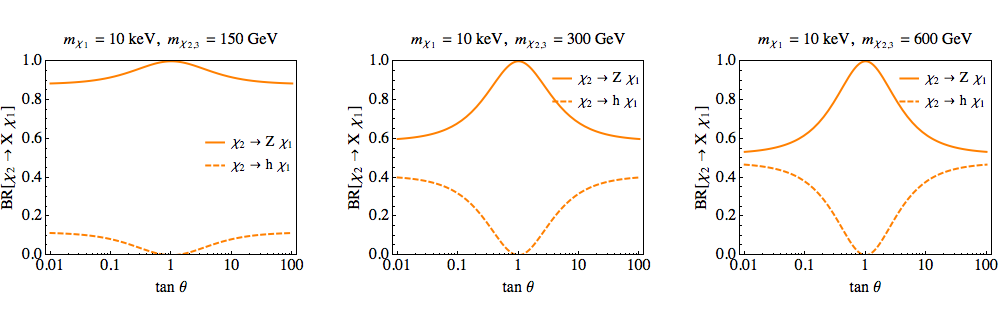}\\
 \vspace{-0.5cm}
  \includegraphics[width=0.91\textwidth]{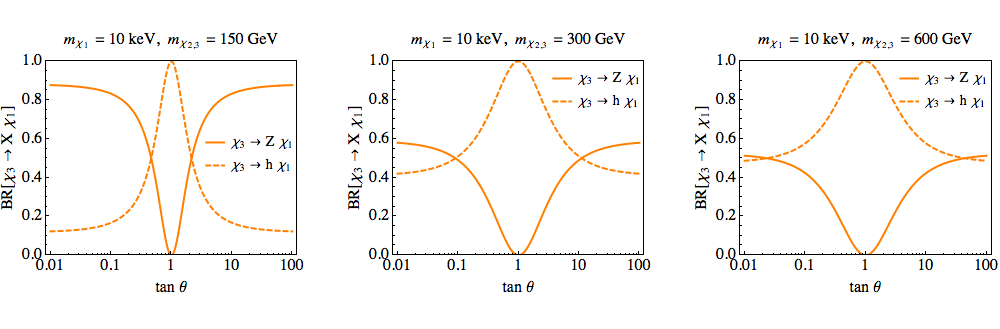}
  \caption{\label{fig:BRs}
  Branching ratios of $\chi_2$ (first row) and $\chi_3$ (second row) as a function of $\tan\theta\equiv y_u/y_d$, for different choices of their mass. If $\tan\theta<0$, $\chi_2$ and $\chi_3$ simply exchange role.}
\end{figure}
\noindent $\bullet$ {\bf Displaced $h$ and/or $Z$ + $\MET$:} Most of the mediator
production modes will eventually produce a pair of heavy neutral
fermions  ($\chi_2 \chi_2$, $\chi_2 \chi_3$ or $\chi_3 \chi_3$),
possibly with extra soft objects that will go undetected. Indeed, as
shown above, the relic abundance requirement
implies that the charged fermions decay dominantly into the heavy neutral
ones, $\chi_{2,3}$, plus soft pions. Given the possible decay modes of $\chi_{2,3}$, 
our key collider signature is thus characterised by a final state with displaced $ZZ$, $hh$ or $Zh$, 
plus missing momentum, as illustrated in Figure~\ref{fig:feyn}. 
By summing all the possible production modes, the process is symmetric in the $\chi_2 \leftrightarrow \chi_3$ exchange, since
$\psi^\pm$ decays democratically into $\chi_{2,3}$, cf.~Eq.~(\ref{eq:psi2pi}).
Hence, the precise signal yield in each of the three channels $ZZ$, $hh$ and $hZ$ is determined by the branching fractions of the two neutral
fermions $\chi_{2,3}$. In Figure~\ref{fig:BRs}, we show the
   branching fractions of the $\chi_{2,3}$ decays into $h\chi_1$
  (dashed line) and $Z\chi_1$ (continuous line) as a function of $\tan \theta$ and for several benchmark masses.
As we can see, for $\tan\theta \approx 1$, i.e.~$y_u\approx y_d$, one of the two heavy fermions decays predominantly into $Z+\chi_1$ and the other one into $h+\chi_1$, independently of their mass. 
This leads to final states with a balanced sample of $hh$ (25\%), $ZZ$ (25\%) and
$hZ$ (50\%). The same is true for $\tan \theta \gg 1$ or $\tan\theta \ll 1$
and when the mass of the neutral fermions is much larger than the
Higgs mass (where effectively one has ${\rm BR}[\chi_{2,3} \to Z
  \chi_1] = {\rm BR}[\chi_{2,3} \to h \chi_1] = 50\%$).  The only
configuration where there is not a balance in $h$ and $Z$ is when the
mass of the neutral fermions is close to the Higgs mass. In this latter
case, kinematics favor the decays into $Z+\chi_1$, and hence final
states with $ZZ+\MET$ are more probable.

 In the next subsection, we will estimate the constraints on the three
 final states with displaced $ZZ+\MET$, $hh+\MET$ or $hZ+\MET$ that can be
 obtained from existing LHC searches at 13 TeV for displaced signatures,
 and we will subsequently study the impact on the parameter space of
 our model.  Notice that searches performed at the LHC with
 $\sqrt{s}$\,=\,8~TeV can also be sensitive to the main signatures of our
 model, $ZZ+\MET$, $hh+\MET$ or $hZ+\MET$.  A number of such
 searches have been considered in Ref.~\cite{Liu:2015bma} and
 reinterpreted in terms of supersymmetric models.  In particular, our
 scenario is similar to the case of Higgsinos decaying into gravitino
 in gauge-mediated SUSY models considered in~\cite{Liu:2015bma}, which
 is constrained mainly by a search for displaced dileptons
 \cite{CMS:2014hka} and a search for displaced jet pairs
 \cite{CMS:2013oea,CMS:2014wda}, both performed by CMS.  In the next
 subsection, we show a comparison of the sensitivity of these searches
 with the 13 TeV analysis we are going to recast. Note that
 our model and the Higgsino-gravitino scenario considered
 in~\cite{Liu:2015bma} differ in an important aspect.  The main
 difference is that the mass splitting between the neutral components
 of the Higgsino multiplets in SUSY scenarios is typically sizeable
 ($\gtrsim {\cal O} (1)$ GeV) due to mixing with gauginos, even if such
 particles are substantially heavier than the Higgsino (up to ${\cal
 O}(10)$ TeV).  As a consequence, the heaviest neutral component of the
 Higgsino decays mostly in the lightest component plus soft SM
 particles and the balance among the $hh$, $hZ$, and $ZZ$ final states
 only depends on the branching ratios of the lightest neutral
 component of the Higgsino into the gravitino plus $h$ or $Z$.  In
 contrast, in the Singlet-Doublet freeze-in model, the mass splitting
 between the heavy neutral components is so small\,---\,as shown by
 the expressions in Eq.~(\ref{eq:mlim})\,---\,that the two neutral
 fermions always decay directly to $\chi_1 Z$ or $\chi_1 h$ with
 branching ratios as illustrated in Figure~\ref{fig:BRs}.  \\[2mm]
\noindent $\bullet$ {\bf {Searches for prompt decays:}}
For small values of the decay length of the mediators (corresponding
to moderate/large values of $y$), we expect that limits from standard
prompt searches can be effective. A combination of recent searches at
the LHC with $\sqrt{s}$\,=\,13 TeV for production of supersymmetric
charginos and neutralinos can be found in
Ref.~\cite{Sirunyan:2018ubx}.
Possible final states are $hh+\MET$, $ZZ+\MET$ and
$hZ+\MET$, which are typical signatures of searches targeting
Higgsino-like neutralinos in gauge mediated supersymmetry breaking
\cite{Matchev:1999ft}, or Higgsinos decaying into light Bino, see
e.g.~\cite{Calibbi:2014lga}.
In these final states, limits to the Higgsino mass up to
600-700 GeV were obtained.\footnote{Searches performed with the
  dataset of the 8 TeV run of the LHC are comparatively much less
  sensitive, constraining Higgsino masses up to around 250 GeV
  \cite{Calibbi:2014lga}, so that we are not going to consider them
  here.}
A second type of relevant final states
are $WZ+\MET$ and $Wh+\MET$, which are possible in our model
for moderate/large values of $y$, such that the charged fermions
$\psi^{\pm}$ decay promptly into $W^{\pm}\chi_1$, cf.~Figure~\ref{fig:BRchargino}.

Note that the configurations of the model giving rise to the observed relic abundance
through freeze-in considered in Sec.~\ref{sec:FI} never
give rise to prompt decays, i.e.~the decay length is always larger
than about 1 mm.
We discuss thus these prompt decay searches briefly.  For more
detailed discussions of the prompt signatures of the Singlet-Doublet
model, see Refs.~\cite{Enberg:2007rp,Calibbi:2015nha}.

\noindent $\bullet$ {\bf {Mono-X searches:}}
In the region with very large decay length, where the neutral fermions escape the detectors, mono-X searches could be the only strategy to look for this model at the LHC 
(besides the disappearing charged
tracks associated to the charged fermion).
In this regime the collider signature of our model is very similar to an Higgsino dark matter scenario in which the mass splitting among the Higgsino components is tiny, as already mentioned.
There have been several investigations on this scenario and the corresponding mono-X signatures, e.g.~in \cite{Schwaller:2013baa,Baer:2014cua,Han:2014kaa,Low:2014cba,Arbey:2015hca,Nelson:2015jza}. 
Some of these investigations have exploited the soft leptons that would be present for a mass splitting of the order of few GeV, which is however not the case of our model. Instead, the case of a pure mono-jet signal has been shown to be not promising, with an estimated
reach on the Higgsino mass of order $200$ GeV at HL-LHC with $3000$ fb$^{-1}$ \cite{Low:2014cba}. 
Hence we decide not to include these signatures in our analysis.

\subsection{Recasting strategy for displaced $h$ and $Z$ +$\MET$}
\label{sec:recast-strat-displ}
\begin{table}[t]
\begin{center}
\begin{tabular}{|c|c|c|c|}
\hline
Experiment & Final state & $\mathcal{L}$ , $\sqrt{s}$ & Ref.\\
\hline
\hline
 \Red{ATLAS} & \Red{DV+$\MET$} & \Red{32.8 fb$^{-1}$ ,  13 TeV} & \cite{Aaboud:2017iio} \\
\hline
ATLAS & lepton-jets & 3.4 fb$^{-1}$ ,  13 TeV & \cite{ATLAS:2016jza} \\
\hline
ATLAS & jets & 3.2 fb$^{-1}$ ,  13 TeV & \cite{ATLAS:2016olj} \\
\hline
CMS & jets & 2.6 fb$^{-1}$ , 13 TeV & \cite{Sirunyan:2017jdo} \\
\hline
CMS & $\mu,e$ & 2.6 fb$^{-1}$ , 13 TeV & \cite{CMS:2016isf} \\
\hline
\hline
\end{tabular}
\end{center}
\caption{Summary of 13 TeV ATLAS and CMS searches for displaced signatures possibly relevant for the final state under study.}
\label{tab:displ}
\end{table}

The aim of this subsection is to estimate the current LHC limit on displaced $ZZ$,
$Zh$ and $hh$ + $\MET$ using public information from ATLAS and CMS searches.
In Table~\ref{tab:displ},
we report on the relevant ATLAS and CMS searches for displaced
signatures, focussing on the most recent analyses at 13 TeV.  Among
these searches, we identify the recent ATLAS analysis on displaced
vertices (DV) with jets and $\MET$ \cite{Aaboud:2017iio} as the most promising for our
scenario. 
The motivation is manifold: 
(i) this analysis exploits the largest available dataset among
those listed in Table~\ref{tab:displ}; 
(ii) the large hadronic branching fractions of $h$ and $Z$
  imply that our model yields a sizeable production cross section in this channel;
(iii) our final states contain a relevant source of $\MET$, and the
  analysis of Ref.~\cite{Aaboud:2017iio} is the only one targeting it with a dedicated selection;
(iv) and finally,  detailed auxiliary material is provided with
  the information needed for a recasting \cite{auxiliary_ATLAS}.

The ATLAS DV+$\MET$ analysis \cite{Aaboud:2017iio}
targets final states with at least one displaced vertex with jets and large missing
transverse momentum.  The results are interpreted in a model with long-lived gluinos decaying into jets and the lightest
neutralino.  In the auxiliary material \cite{auxiliary_ATLAS}, the
efficiencies for the missing momentum and the displaced vertex
reconstruction are provided. In
particular, the efficiencies of the displaced vertex reconstruction are given prior to detector simulation, as a
function of the invariant mass of the vertex, of the number of tracks
and of the displacement.

In order to estimate the sensitivity of this search on our final states, we
have first implemented the model in {\tt FeynRules}~\cite{Alloul:2013bka}, and then
simulated the relevant samples with {\tt MadGraph5}~\cite{Alwall:2014hca},  combined with {\tt
  Pythia8}~\cite{Sjostrand:2014zea} for the parton showering and the underlying $pp$ collision, and {\tt Delphes3}~\cite{deFavereau:2013fsa} (with the standard ATLAS card)  for the detector simulation.
 The displacement is applied to the simulated events a posteriori, taking into account the four
momenta of the long-lived particle in order to properly compute the displacement,\footnote{In this approach we neglected possible distortions of the kinematic distributions of the final state charged tracks due to the displacement.}
which is obtained by sampling an exponential distribution with mean decay length $c \tau_{\chi_{2,3}}$.
In Appendix \ref{app:reca}, we discuss the details of the
selection and the validation of our implementation. The latter was  performed
by reproducing the exclusion limits set by the ATLAS search on the
simplified model they considered with long-lived gluino.

After this validation, we can now estimate the efficiency of the ATLAS DV+$\MET$ analysis in the final states we are interested in,
which are $ZZ+\MET$, $hh+\MET$ or $hZ+\MET$.\footnote{If $Z$ or $h$ decay into $b\bar{b}$, (some of) the resulting tracks may have an additional displacement, which makes the reconstruction of the DV more involved, as discussed in more detail in Appendix \ref{app:reca}. In the following, we neglect possible issues related to this for the reasons discussed in the appendix.} 
We do this as a function of the lifetime $\tau_{\chi_{2,3}}$ and of the mass of the long-lived particles,
which are simply the two neutral fermions $\chi_2$ and $\chi_3$ produced in pairs
(cf.~Appendix \ref{app:reca} for plots displaying the resulting efficiencies). The mass 
is important in order to determine the boost factor in the
displacement, as well as to get the correct $p_T$ distribution of the
displaced tracks. 
We can now use the obtained selection efficiencies to evaluate the reach of the
ATLAS analysis in three simplified models with fixed branching fractions, that serve for illustrative purposes:
\begin{itemize}
\item[i)] ${\rm BR}[\chi_{2,3} \to h \chi_1] = 100\%$;
\item[ii)] ${\rm BR} [\chi_{2,3} \to Z \chi_1] = {\rm BR}[\chi_{2,3} \to h \chi_1] = 50\%$;
\item[iii)] ${\rm BR}[\chi_{2,3} \to Z \chi_1] = 100\%$.
\end{itemize}

\begin{figure}[t]
  \centering
 \includegraphics[width=0.55\textwidth]{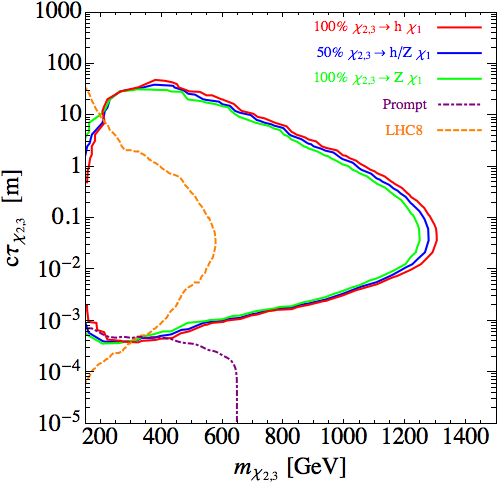}
 \caption{\label{fig:LHC_bound}
Estimated exclusion curves from collider searches in the plane of the
decay length versus the mass of the heavy neutral fermions. Our recasting of the ATLAS DV+$\MET$ search,
associated to the final states of $hh+\MET$, $hZ+\MET$ or $ZZ+\MET$,
for the simplified models i), ii) and iii) (see the text for details) are shown with red, blue and green
solid lines respectively. The orange dashed line is the exclusion of
displaced searches at 8 TeV LHC, as estimated in
  \cite{Liu:2015bma}. The purple dashed-dotted line is our estimate
of the impact of the prompt searches at 13 TeV performed by CMS
\cite{Sirunyan:2018ubx}. }
\end{figure}

In order to constrain the above simplified models, 
we consider the total production cross section of the doublet fermions states,
 computed at NLO by {\tt Prospino2} \cite{Beenakker:1999xh}, 
 summing all production modes shown in (\ref{eq:modes}),
 corresponding to the solid red line in Figure~~\ref{fig:xsec}.
With no background in the signal region,
the parameter configuration of a model is excluded at 95\% confidence level (CL) or more if it yields a number of selected events $\ge 3.0$.
 The resulting estimated exclusion is depicted in Figure~\ref{fig:LHC_bound} by the three solid lines.     
 As we can see, the difference in the efficiencies among the three simplified models results in only a small impact on the sensitivity.
Moreover, the largest doublet mass (about 1.3 TeV) is probed for
a decay length around $c\tau \approx 5$ cm. Also, the exclusion curves
 are not symmetric in $c\tau$ with respect to this maximal reach.
 This is due to the fact that the exponential distribution determining the displacement is 
 falling very rapidly for a displacement larger than a given $c\tau$, while it goes to zero
 less steeply for displacement smaller than $c\tau$.  This also
 explains why the reach of the analysis extends to regions with very large decay lengths, up to  $c\tau \approx 50$ m.

In Figure~\ref{fig:LHC_bound}, we also show for comparison the exclusion from the $8$
TeV searches, as reported
by Ref.~\cite{Liu:2015bma}.  This is depicted as a dashed
orange line and includes both searches targeting displaced leptons
\cite{CMS:2014hka} and displaced di-jets
\cite{CMS:2013oea,CMS:2014wda}.  The displayed 8 TeV limit has been
obtained in Ref.~\cite{Liu:2015bma} in a simplified model with an
Higgsino-like neutralino undergoing displaced decays into gravitino
plus $Z$ or $h$ in the large $\tan \beta$ regime, which roughly
corresponds to our simplified model ii).\footnote{The other simplified models
  considered in Ref.~\cite{Liu:2015bma}, corresponding to our cases i)
  and iii), give very similar exclusion power.}  As we can see, in the
region of low doublet mass, the sensitivity of the ATLAS DV+$\MET$
analysis is diminished because the spectrum is compressed and jet/$\MET$ cuts become more severe. 
This is where the 8 TeV searches, in particular the one targeting displaced
dileptons, become instead more efficient, despite the small
leptonic branching fractions of the bosons.\footnote{We also remark that,
  as discussed in detail in the Appendix \ref{app:reca}, our
  implementation of the ATLAS DV+$\MET$ analysis tends to overestimate
  the exclusion in the compressed region (for mass splittings
  $\lesssim 100$ GeV). The complementarity with the 8 TeV searches is
  thus welcome.}
Finally, we also display on the same plot an estimate of the reach of the prompt searches (as a purple dot-dashed line), considering the ${\rm BR}[\chi_{2,3} \to Z \chi_1] = {\rm BR}[\chi_{2,3} \to h \chi_1] = 50\%$ case reported in \cite{Sirunyan:2018ubx}.
We stress that this limit will never be relevant in the parameter region leading to the correct freeze-in dark matter abundance, but we report it here for illustrative purposes. In order to draw this line, we have compared the cross section limits reported in Ref.~\cite{Sirunyan:2018ubx} to the total production cross section of the doublets multiplied by the probability that both produced particles decay promptly given a certain mean decay length $c \tau_{\chi_{2,3}}$.\footnote{As a rough estimate, we consider to be prompt the events with a total displacement $\leq 0.5$ mm that we compute based on $c \tau$ only, without taking into account the boost factor.}

\subsection{DV+$\MET$ constraints on the Singlet-Doublet model}
We can now use the recasting presented above to provide estimates for
the ATLAS exclusion on the parameter space of the Singlet-Doublet
freeze-in model.  For this purpose, at each point of the parameter
space, we sum the production cross sections over the 
production channels weighted by the appropriate branching fraction in
order to determine the signal strength for each of the possible final
states.  For instance, the signal cross section in $hh+\MET$ is given by
 \begin{align} 
 \sigma( p p \to h h\chi_1\chi_1) ~= ~ & \sigma(p p \to \psi^+ \psi^-)\times
\text{BR}[\psi^{\pm} \to \pi^\pm\chi_{2,3} ]^2 \times \text{BR}[\chi_{2,3} \to h \chi_1]^2  +\nonumber \\ 
&   \sigma(p p \to \psi^{\pm} \chi_{2,3})
\times \text{BR}[\psi^{\pm} \to \pi^\pm\chi_{2,3} ] \times \text{BR}[\chi_{2,3} \to h  \chi_1]^2 + \nonumber \\ 
& \sigma(p p \to \chi_{2} \chi_{3}) \times \text{BR}[\chi_{2} \to h \chi_1] \times \text{BR}[\chi_{3} \to h \chi_1], 
\end{align}
and analogous expressions can be written for the $hZ+\MET$ and $ZZ+\MET$.  
The production cross section in each channel is hence a function of the parameter
space of the model through the branching fraction dependence on $(y_u,
y_d, \mu, m_s)$. We multiply these three type of signal cross sections
with the corresponding efficiencies (derived in Appendix
\ref{app:reca}) to obtain the final estimate on the number of expected
events.  Each efficiency is also a function of the parameters $(y_u,
y_d, \mu, m_s)$, since it depends on the mass of the long-lived
particle, which is simply $\mu$, and on the decay lengths that follow from Eqs.~(\ref{decay_neutr}-\ref{decay_neutr4}).
For simplicity, we take the average of the decay length of $\chi_2$ and $\chi_3$ as the mean decay length setting the displacement. 
As we have discussed above, 
this is an excellent approximation as long as $\tan\theta \gg 1$ or $\mu\gtrsim 300$ GeV (see Figure~\ref{fig:decaylength}).  
We neglect the extra displacement induced by the decay of the charged
fermion. 
Note that this is indeed typically a small fraction of the
overall displacement in the relevant portion of the parameter space,
as illustrated by the green contours in the right panel of
Figure~\ref{fig:chidecay}.  In our estimate we also consider the same
efficiency in the case in which the neutral heavy fermions are directly
pair produced as in the case in which the neutral fermions are
produced through the decay of the charged fermion.  We checked this
hypothesis on a few benchmark points and it induces an effect of at most
$20\%$, which is largely negligible for the purpose of our recasting. As for the case of the simplified
models considered in Sec.~\ref{sec:recast-strat-displ}, we employ the NLO cross sections computed by {\tt Prospino2} \cite{Beenakker:1999xh}
and we calculate 95\% CL cross section upper exclusion limits assuming no background.
\begin{figure}[t!]
  \centering
 \includegraphics[width=0.55\textwidth]{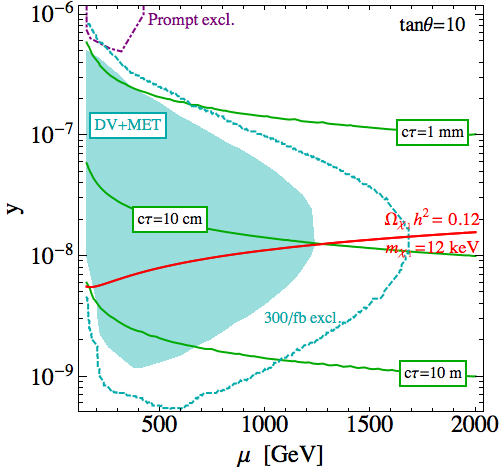}
 \caption{\label{fig:LHC_bound_yplot}
Exclusion capability of the 13 TeV ATLAS search for displaced jets + $\MET$ for the Singlet-Doublet freeze-in model (cyan region labelled as ``DV+MET'') on the plane $(\mu,~y)$ with $m_{\chi_{2,3}}=m_\psi =\mu$.
The dashed cyan line represents the prospected bound with  $300\,\rm fb^{-1}$.
The dot-dashed purple line is our estimate of the limit set by LHC searches for the prompt signature $WZ+\MET$ (see the text for details).
The green contours indicate the average decay length of $\chi_2$ and $\chi_3$. The red line corresponds to the correct relic abundance for $m_{\chi_1}=12$ keV. }
\end{figure}

Under the above assumptions, we can assess the current limits on the
Singlet-Doublet parameter space from the ATLAS DV+$\MET$ search. The
 region excluded according to our recasting is shown with filled cyan colour in Figure
\ref{fig:LHC_bound_yplot}.  Its shape follows
from combining the excluded regions for the simplified models reported in
Figure~\ref{fig:LHC_bound} with the iso-contours of the average
$c\tau_{\chi_{2,3}}$ (denoted as $c\tau$). The latter are shown with green continuous lines
in Figure \ref{fig:LHC_bound_yplot} while the dashed cyan curve gives
the estimated reach of an analogous DV+$\MET$ search with a dataset of
$300$ fb$^{-1}$.\footnote{The estimated curve for the DV+$\MET$ search with a
  dataset of $300$ fb$^{-1}$ results from simply rescaling the
  luminosity and assuming the signal to remain background
  free. While this is an optimistic assumption, it may not be unthinkable that backgrounds can continue to be suppressed at the cost of only a small signal inefficiency.} 
 The red continuous curve shows the ($y$,
$\mu$) combinations that account for all the DM for a 12 keV DM
candidate.
Going above the red
line, i.e. to larger values of the coupling $y$, induces an
overabundant dark matter population, while below the red line it is
underabundant; see Eq.~(\ref{eq:simpleOm}).
Finally, the dot-dashed line delimits the region excluded by LHC
searches for the prompt signature $W Z + \MET$.  

It is remarkable that the sensitivity of the ATLAS DV+$\MET$
search extends to heavy electroweak states and to quite large values
of the decay lengths. This is related to the almost background-free
nature of displaced vertices signatures which renders the search very
efficient even for small signal cross section. Note that the largest
excluded mediator mass is about 1.2 TeV, somewhat smaller than for the
simplified model analysis of Figure~\ref{fig:LHC_bound}.  The reason is as
follows. In the high mass region, when the lifetime, or equivalently
the coupling $y$, maximises the experimental sensitivity ($y \approx
10^{-8}$) the branching fraction of the process $\psi^\pm \to \chi_1
W^{\pm}$ is not completely negligible (up to about $10\%$, see Figure
\ref{fig:BRchargino} left), and hence the signal yield into long-lived
neutral fermion pairs is slightly diminished. In the case of the DV+$\MET$
analysis extrapolation to 300 fb$^{-1}$, we expect to  probe masses of the neutral fermions
up to $1.7$ TeV and decay lengths as large as $100$ m.

Let us add here a remark on the uncertainties on our recasting and their effects on the estimated limits in Figure~\ref{fig:LHC_bound_yplot}.
Given the steep fall of the production cross section as a function of the mediators' mass (see Figure \ref{fig:xsec}), we expect that even $\ord{1}$ modifications of our estimated efficiencies would have a small impact on the mass reach (for instance a 50\% change in the efficiency would only correspond to a change of around 10\% in the mediators' mass limits).\footnote{Note in particular that this applies to the possible issues associated with $b$-jets, discussed at the end of Appendix \ref{app:reca}, that would at most reduce the signal strength by $\approx$ 25\%.}

Let us stress that the collider bounds presented in
Figure~\ref{fig:LHC_bound_yplot} are expected to be independent of the
dark matter mass for $m_{\chi_1}$ below the GeV scale. The only curve
that is affected by the $m_{\chi_1}$ parameter is the relic abundance
continuous red contour. 
Considering larger
values of the dark matter mass the red line would be shifted to lower
values of $y$. 
As a result, the dark matter candidates with $m_\chi>12$ keV
(i.e. compatible with the Lyman-$\alpha$ bound discussed in
Sec.~\ref{sec:cosmo-bounds}) are concerned with the excluded region
below the red curve of Figure~\ref{fig:LHC_bound_yplot}.
For instance, from the right panel of Figure \ref{fig:Omega},
one can deduce that for e.g.  $m_{\chi_1} \approx 1$ MeV the $\Omega
h^2= 0.12$ contour should appear at $y \approx 10^{-9}$ in Figure~\ref{fig:LHC_bound_yplot}. This
corresponds to larger values of $c \tau$ where the DV$+\MET$ search
loses sensitivity in such a way that no constraint can be set at
present. 

For completeness, let us briefly discuss the prompt decay
constraints. In the upper part of Figure \ref{fig:LHC_bound_yplot},
the size of the coupling $y$ is such that the mediators decays are
prompt. In particular, the charged fermion $\psi^{\pm}$ predominantly
decays into $W^{\pm} \chi_1$.\footnote{The other prompt signatures
  discussed above  are less
  sensitive as the production cross section is sensibly lower for
  $\chi_2,\chi_3$ production only.  The latter is indeed almost one
  order of magnitude smaller than the total doublet production as seen
  in Figure \ref{fig:xsec}.} In order to estimate the corresponding
constraint, we have computed the $WZ+\MET$ production cross section in
the Singlet-Doublet model multiplied by the probability that both
heavy particles decay promptly, using the same approximations as for
the prompt exclusion in Figure \ref{fig:LHC_bound}. Comparing the
latter results with the limits on the cross section given in
Ref.~\cite{Sirunyan:2018ubx} we exclude the region delimited by the
dot dashed purple line of Figure \ref{fig:LHC_bound_yplot}. As
discussed above, such constraint lies however in a zone of the
parameter space where the frozen-in dark matter scenarios with masses
above the Lyman-$\alpha$ bound give an overabundant dark matter relic
density.

\section{Displaced vertices vs Cosmology for freeze-in DM}
\label{sec:discussion}
\begin{figure}[t]
  \centering
 \includegraphics[width=0.55\textwidth]{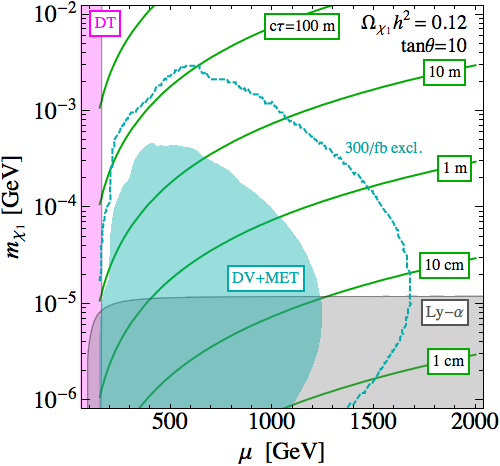}
\caption{\label{fig:combi_plot}
Combined constraints on the mediator mass vs DM mass plane ($m_{\chi_{2,3}}=m_\psi =\mu$).
Our estimate of the ATLAS DV + $\MET$ exclusion is shaded in cyan (``DV+MET''), the magenta region is excluded by disappearing tracks (``DT''), the Lyman-$\alpha$ bound is shown in gray (``Ly-$\alpha$''). Green contours correspond to the average $\chi_{2,3}$ decay length. The cyan dashed line is the estimated exclusion of LHC with 300 fb$^{-1}$.
The coupling $y$ is fixed such that $\Omega_{\chi_1}h^2=0.12$ everywhere.}
\end{figure}

We can now combine the LHC limits and the cosmological bound derived
in the previous sections, in order to characterise the experimental
sensitivity on the viable parameter space of the freeze-in
Singlet-Doublet model.  As at the end of Section \ref{sec:FI}, we
present our results in the DM mass vs mediator mass plane fixing in
each point the coupling $y$ to the value that accounts for the
observed relic abundance through the freeze-in mechanism.  On the same
two dimensional plane, we can show the combination of the existing
(and future) constraints on the model.  Our summary plot is shown in
Figure \ref{fig:combi_plot}.  As before, the green lines indicate
contours of fixed average decay length of the neutral fermions, which controls the phenomenology at colliders.  
The magenta shading at low mediator masses represents the region excluded
by searches for disappearing charged tracks
(DT)~\cite{Atlas-DT,Atlas-DT2}.  It does not depend on the DM mass (or
equivalently on the value of the $y$ coupling) since, in this region,
the decay length of the charged fermion is independent of $m_{\chi_1}$,
as can be seen in the right panel of Figure~\ref{fig:chidecay}.  The
cyan region and the dashed cyan line are the estimated exclusion and future
prospect of the ATLAS DV+$\MET$ search,
discussed in Section~\ref{sec:LHC}.  The gray region, finally, is excluded by
the Lyman-$\alpha$ forest data, as discussed in Section~\ref{sec:cosmo-bounds}.

Figure~\ref{fig:combi_plot} summarises the findings of this paper, 
as it nicely shows the interplay between collider searches for displaced signatures and cosmological constraints
in our freeze-in dark matter model.
On the one side, the observed DM abundance implies a relation among the parameters of the theory, 
leaving only two free parameters (plus a third one, $\tan\theta$, that affects the phenomenology of the model very mildly in our limit,
as we discussed in the previous sections).
On the other side, for the range of decay lengths that are a priori optimal for studying displaced signatures at the LHC
($\mathcal{O}(10)$ cm) and $\mu$ scales within the reach of the
collider ($\mu \lesssim \mathcal{O}(1)$ TeV), our dark matter model can leave
a testable imprint on small scale structures.  In such a region, we have a
complementary constraint from the Lyman-$\alpha$ forest observations, 
which is essentially independent of the mediator
mass. In contrast, the reach of LHC searches is
intrinsically limited by the production cross section of the mediators,
hence by their mass, and by the size of the detector. Due to
the very low background of the recast search and the large dataset
available, the current LHC limit actually extends to rather large values of
the mediators' lifetime and, likewise, it can probe DM
masses larger than those tested by cosmology, reaching up to
$m_{\chi_1} = \mathcal{O}(1)$ MeV.

\section{Summary and Conclusions}
\label{sec:conclusions}
Despite many experimental and theoretical efforts, the nature of dark
matter remains a mystery.  It is thus timely to look for DM beyond the
most popular paradigms. In this work, we considered the case of a dark
matter candidate with such a tiny coupling that it never reaches
thermal equilibrium with the SM in the early universe.  It is well
known that, despite such suppressed interactions, the observed dark
matter density can be accounted for by the freeze-in mechanism, with
the dark matter being produced, for instance, via the decay of
thermalised mediators. Within this context, we studied the case of the
Singlet-Doublet dark matter model that consists in extending the SM
with a pair of Weyl electroweak doublet fermions and a singlet
Majorana fermion. The new fermions interact with the SM through gauge
interactions and/or the Higgs portal induced by Yukawa interaction
terms that couple the doublet and the singlet fermions to the Higgs
particle.  The Singlet-Doublet model rests thus on 4 free parameters
only: 2 new mass scales (the doublet mass scale $\mu$ and the singlet
mass scale $m_s$), and two Yukawa couplings ($y_u$ and $y_d$).  We
have shown that, considering these couplings in the range $ [10^{-8},
  10^{-10}]$ together with a doublet mass scale $\mu$ larger than the
Higgs mass, the lightest neutral fermion, which is essentially the
singlet Majorana fermion, can account for the whole dark matter
abundance via the freeze-in mechanism.  In this regime, the DM is
light with a mass between a few keV up to hundreds of MeV.

 Such a dark matter
scenario could seem hopelessly beyond the reach of any dark matter
experimental search. We show instead that the range of model
parameters required for a successful freeze-in naturally gives rise to
long-lived/displaced collider signatures that are already strongly bounded by
the present LHC data.
 In addition,
it is well know that thermal warm dark matter candidates of a few keV
are also constrained by cosmology due to their free-streaming
suppressing the growth of small scale structure.  Even though
frozen-in dark matter was never in thermal equilibrium in the early
universe, Lyman-$\alpha$ bounds turn out to constrain the dark matter to be always
heavier than 12 keV. In the low dark matter mass region, the
model features thus both exotic LHC signals and a testable imprint on
cosmology providing two complementary handles to probe the same scenario.

Concerning the collider searches, 
the relevant
signatures 
of this model consist of
disappearing charged tracks, related to the production of the charged component of
the doublet $\psi^\pm$, and displaced $h$ and/or $Z+ \MET$, associated
to the decay of the two neutral fermions $\chi_{2,3}$. In the first
case, $\psi^\pm$ decays with a small displacement (about 1 cm)
and our scenario is essentially the case of pure Higgsino DM in supersymmetric
models.
Current searches for disappearing tracks thus constrain the doublet mass scale $\mu$ to be larger
than $150$ GeV. 
In the second case, the $\chi_{2,3}$ fermions decay with
displacements in a wide range, from centimetres to kilometres,
depending on the point of the viable parameter space of interest
(i.e.~for $y_u,\,y_d$ giving rise to the right relic abundance through
freeze-in). We have argued that, at present, the most constraining search was
provided by ATLAS in Ref.~\cite{Aaboud:2017iio} and we have reinterpreted
its results in the framework of the Singlet-Doublet dark matter
model. 
According to our recasting, this analysis can exclude scenarios with a decay
length of the heavy neutral mediators as large as $\sim 50$ m, 
mediator masses as large as 1.2 TeV,
and dark
matter candidates with masses as large as 500 keV.

In Figure~\ref{fig:combi_plot}, we have brought together all the experimental signatures
which can probe
the viable parameter space where the
freeze-in production mechanism gives rise to the correct dark
matter abundance. This nicely illustrates the interplay between collider
searches and cosmology for frozen-in dark matter.

An interesting extension of our work is to enlarge the experimental reach on the parameter space of the model.
The LHC sensitivity, shown in Figure~\ref{fig:combi_plot}, could be
improved towards large mediator masses, or towards small or large
decay lengths. This is possible on all these three fronts by
exploiting the presence of a displaced $Z$ or $h$ resonance, both in
hadronic and leptonic decay channels, such that some of the event
selection requirements that currently limit selection efficiencies can
be relaxed, while keeping backgrounds to a negligible level. Also, at
higher luminosities a dedicated event selection would help to suppress
the increasing backgrounds. As a result of our study, we thus advocate
dedicated experimental searches for displaced $Z+\MET$ or $h+\MET$
signatures, potentially in association with an extra identified $Z$ or
$h$ boson. On the other hand, it would be interesting to also probe
the case with large/moderate dark matter mass and very long-lived
mediators (upper part of Figure~\ref{fig:combi_plot}).  For this
purpose, one could for instance estimate the reach of the proposed
detector MATHUSLA \cite{Curtin:2017izq} on this scenario.

Finally, we stress again that 
interplay between exotic collider signatures and cosmology constraints
go beyond the Singlet-Doublet model and apply to a large class of simplified models of freeze-in dark matter where the production occurs through decays of heavy mediators in thermal equilibrium with the SM bath.
From the model building perspective, it would be interesting to 
investigate such complementarity in other models,
also including those where the freeze-in is not realised through the decays
of heavy mediators, but via scattering processes and/or via
non-renormalisable interactions \cite{Elahi:2014fsa}. We leave these
interesting possibilities for future works.

\subsection*{Acknowledgments}
We are grateful to Francesco D'Eramo, Michele Frigerio, Julian Heeck, Lawrence Lee, Daniele Teresi, Pantelis Tziveloglou, and Bryan Zaldivar for
useful discussions.  LLH is supported by the Fonds National de la Recherche Scientifique.  AM and LLH are supported by
the Strategic Research Program \textit{High-Energy Physics} and the Research Council of the Vrije Universiteit Brussel.  AM and SL are
supported by FWO under the ``Excellence of Science - EOS'' - be.h project n.30820817.

\appendix

\section{Yukawa interactions and mixing matrix}
\label{app:masses}
\subsection{Yukawa interactions}
For the contraction of the indeces in the Lagrangian of Eq.~(\ref{eq:lagr}), we follow the conventions of Ref.~\cite{Calibbi:2015nha,Lopez-Honorez:2017ora}. 
In particular, the Yukawa interactions
\begin{equation}
\label{eq:Yuk}
- y_d ~\psi_d \cdot H \,\psi_s - y_u  ~H^{\dagger} \psi_u \,\psi_s +h.c.
\end{equation}
can be explicitly written as
\begin{equation}
  - y_d ~\psi_{d\,i}  H_j \epsilon^{ij}\,\psi_s-y_u~\psi_{u\,i} H ^{*\,i}\,\psi_s+h.c.\,,
\end{equation}
where $i$ and $j$ are $SU(2)_L$ indices.

\subsection{Approximate expression of the rotation matrix}
In the limit (\ref{eq:lim}), the mass eigenvalues at the first order
in the couplings $y_{u,d}^2$ result as shown in Eq.~(\ref{eq:mlim}).
We report here the rotation matrix, defined as in Eq.~(\ref{eq:Udef}),
at leading order in $y_u$ and $y_d$:
\begin{equation}
U = 
\left(
\begin{array}{ccc}
1 & -\frac{v}{2 \sqrt{2}} \left( \frac{y_u-y_d}{\mu +m_s}+\frac{y_u+y_d}{\mu -m_s}\right) & 
\frac{v}{2 \sqrt{2}} \left( \frac{y_u-y_d}{\mu +m_s}-\frac{y_u+y_d}{\mu -m_s}\right) \\
\frac{v}{2} \frac{y_u-y_d}{\mu +m_s}& -\frac{1}{\sqrt{2}} & \frac{1}{\sqrt{2}} \\
-\frac{v}{2} \frac{y_u+y_d}{\mu -m_s}& -\frac{1}{\sqrt{2}} & -\frac{1}{\sqrt{2}} \\
\end{array}
\right).
\end{equation}
We omit the $\mathcal{O}(y^2_{u,d})$ terms that are needed in order to diagonalise correctly the mass matrix $\mathcal{M}$ obtaining the eigenvalues shown in Eq.~(\ref{eq:mlim}). In fact, in our parameter regime, the above expression
suffices to reproduce the rotations resulting from numerical diagonalisation to high accuracy. Hence, we employ it
to derive the expressions for the decay widths of the heavy particles reported in Section \ref{sec:decays}.

\section{Recasting of the ATLAS search}
\label{app:reca}
In this Appendix, we provide details about the recasting of the ATLAS search of Ref.~\cite{Aaboud:2017iio} that we employed in order to set limits on the Singlet-Doublet model and in particular on displaced neutral bosons + $\MET$ final states. 

The signature that we consider both in the validation (gluino-neutralino simplified model) and in the Singlet-Doublet model 
is constituted by a pair of heavy long-lived particles decaying into charged tracks plus missing energy.
For the case of two neutral heavy fermions, the process is depicted in Figure \ref{fig:feyn}, with the long-lived particle highlighted in red.

We first review the selection cuts of the search, we then validate our simulation with the simplified model studied in the ATLAS analysis, and then we apply the same recasting to our dark matter model.

\paragraph{Selection criteria of the ATLAS DV+$\MET$ search \cite{Aaboud:2017iio}.}
The ATLAS analysis \cite{Aaboud:2017iio} is explained in detail in the auxiliary material in \cite{auxiliary_ATLAS}. 
The search targets displaced vertices and missing transverse momentum. 

The displaced vertices are identified by analysing the associated displaced tracks. First, a selected displaced track should satisfy the following requirements:
\begin{itemize}
\item The track is associated to a stable particle;
\item The particle has a transverse momenta $p_T>1$GeV;
\item The transverse impact parameter $d_0 \equiv   R_{decay} \sin \Delta \phi > 2$mm, where $R_{decay}$ is the transverse decay length and $\Delta \phi$ is the azimuthal angle between the heavy decaying particle momentum and the track momentum.
\end{itemize}
With the following selected tracks, one can construct a candidate displaced vertex which should satisfy the following criteria:
\begin{itemize}
\item The transverse displacement $R_{decay}$ should be within $4$ and $300$ mm;
\item The longitudinal displacement should be smaller than $|z_{decay}| < 300$ mm;
\item The number of associated charged tracks should be $n_{tracks} \geq 5$;
\item The invariant mass of the vertex should be larger than $10$ GeV (a pion mass for the tracks is assumed).
\end{itemize}

Given the previous strategy to select displaced tracks and displaced vertices, events are hence required to satisfy the following conditions.
\begin{enumerate}
\item $\MET^{\text{truth}}> 200$ GeV where $\MET^{\text{truth}}$ is the missing energy at truth level, here interpreted as the magnitude of the transverse component of the vector sum of the
dark matter momenta.
\item On $3/4$ of the events, the ATLAS analysis also demands the presence of either 
\begin{itemize}
\item One jet with $p_T > 70$ GeV
\item Two jets with $p_T > 25$ GeV
\end{itemize}
These jets should satisfy the requirement that the scalar sum of the $p_T$ of the charged particles that are not displaced (according to the previous selection) should not exceed $5$ GeV.
\item The events must contain at least one displaced vertex which has passed the selection.
\end{enumerate}

\paragraph{Recasting and validation.} On the selected events, one can then apply the efficiency as reported in the auxiliary material. 
Indeed, the ATLAS collaboration provides the efficiency for reconstructing the displaced vertices as a function of the number of displaced tracks and of their invariant mass. They also provides the efficiency tables as a function of the missing energy at truth level. 
\begin{figure}[t]
  \centering
 \includegraphics[width=0.45\textwidth]{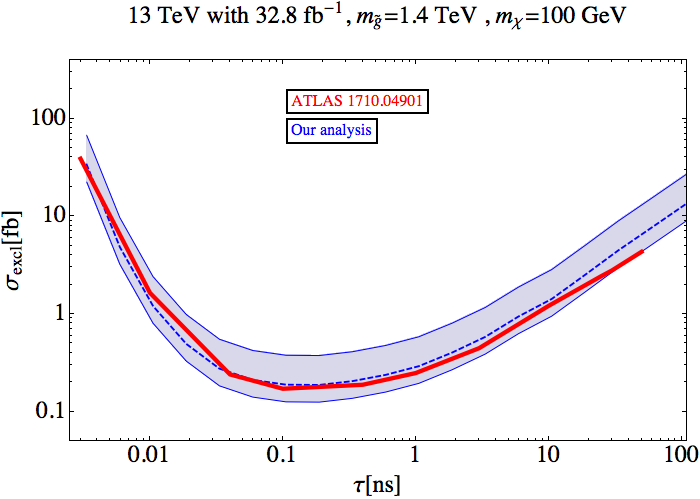}\hspace{5mm}
 \includegraphics[width=0.45\textwidth]{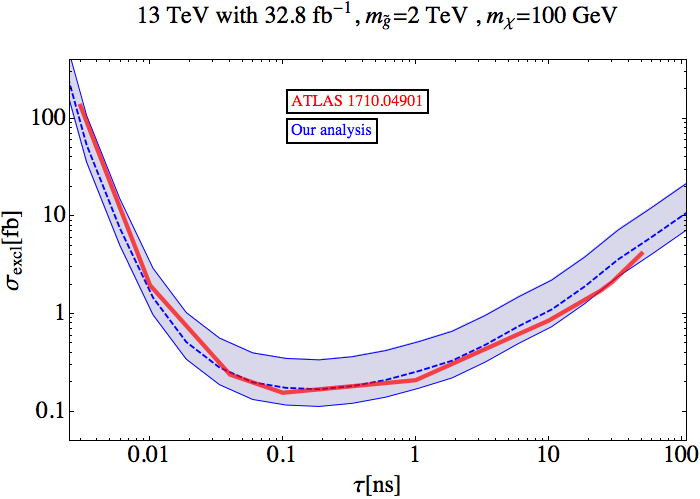} 
 \\
 \includegraphics[width=0.45\textwidth]{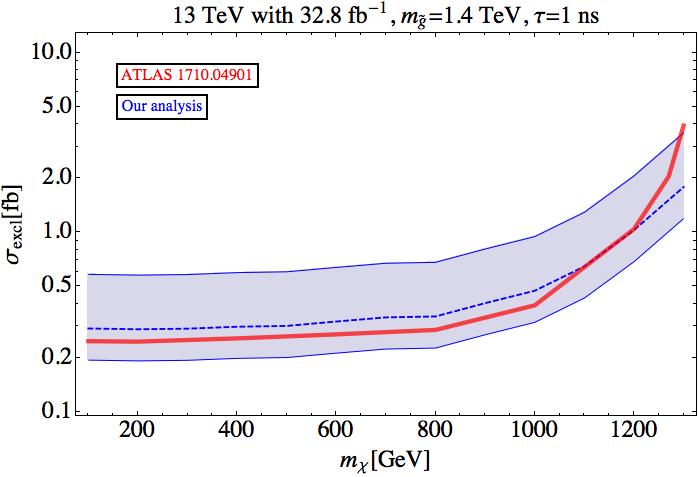}\hspace{5mm}
 \includegraphics[width=0.45\textwidth]{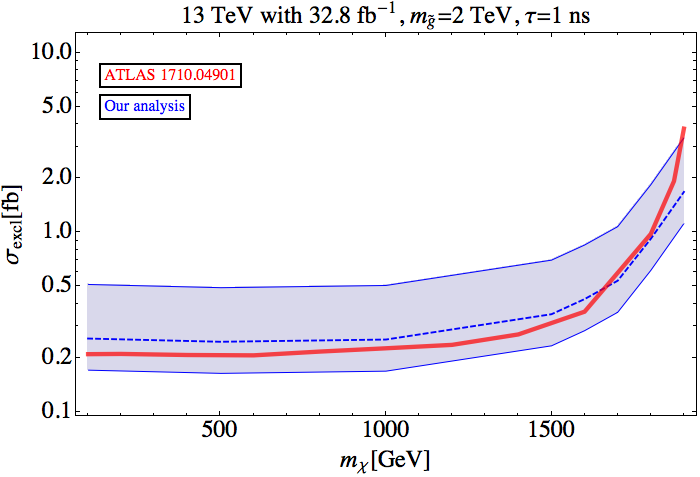} 
  \\
\caption{\label{fig:valid}
Comparison between the result of our simulation and the excluded cross section reported in the ATLAS paper \cite{Aaboud:2017iio} for a long-lived gluino simplified model. The ATLAS results are shown as red lines, while our analysis corresponds to the blue bands. 
In order to  draw our bands, we considered variation of the efficiency of $\pm50\%$.
The upper plots show the excluded cross section as a function of $c\tau$ for a fixed neutralino mass $m_{\chi}= 100$ GeV and two benchmark values for the gluino mass, $m_{\tilde g}=1400,\,2000$ GeV. 
The lower plots show the excluded cross section as a function of $m_{\chi}$ fixing $\tau = 1\,$ ns and the same two benchmarks for $m_{\tilde g}$. }
\end{figure}

In order to recast this analysis, we have simulated LO events of the new physics process with {\tt MadGraph5} + {\tt Pythia8} + {\tt Delphes3} with standard minimal cuts, default parameters for MC and detector simulator (we used the default ATLAS card),
assuming prompt decays of the heavy pair-produced particles. 
We employed generator level information in order to extract the momenta of the two heavy particles and of their associated tracks and in order to introduce the displacement by hand, including the boost factor of the heavy decaying particle. The decay time was generated through an exponential distribution with a mean lifetime $\tau$. With this information we derived the impact parameter of each track and the other relevant geometrical properties. Then we processed the output following the ATLAS selection cuts strategy, including the reconstruction efficiencies.
We first applied this procedure to a simplified model analogous to the one considered in the ATLAS paper, by considering gluino pair production followed by displaced decays into $q \bar q$ plus neutralino.
\begin{figure}[t!]
  \centering
 \includegraphics[width=0.45\textwidth]{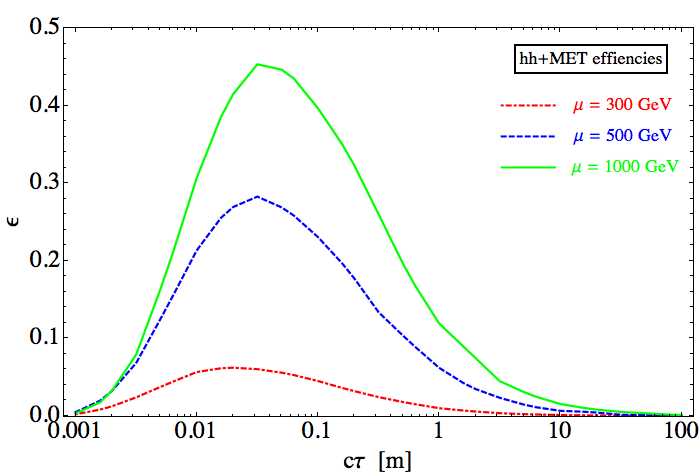} \hspace{5mm}
 \includegraphics[width=0.45\textwidth]{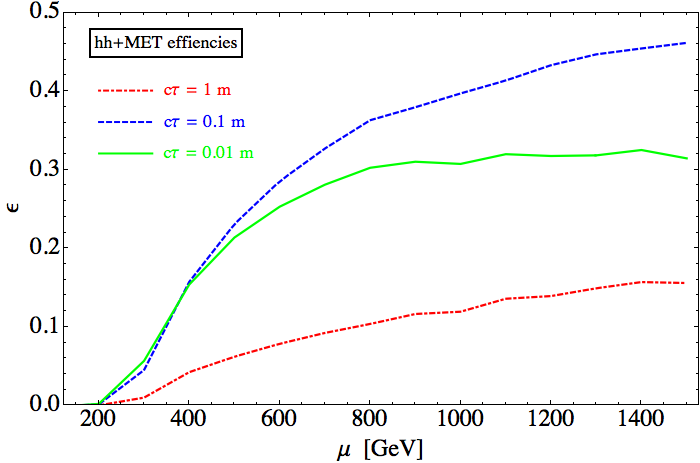} 
 \\
 \includegraphics[width=0.45\textwidth]{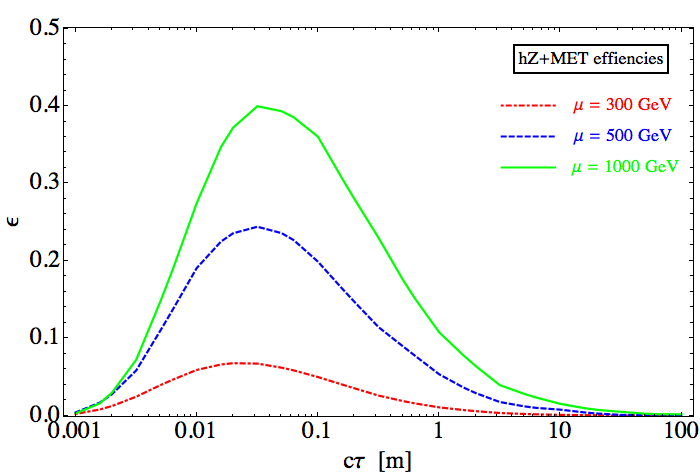}\hspace{5mm}
 \includegraphics[width=0.45\textwidth]{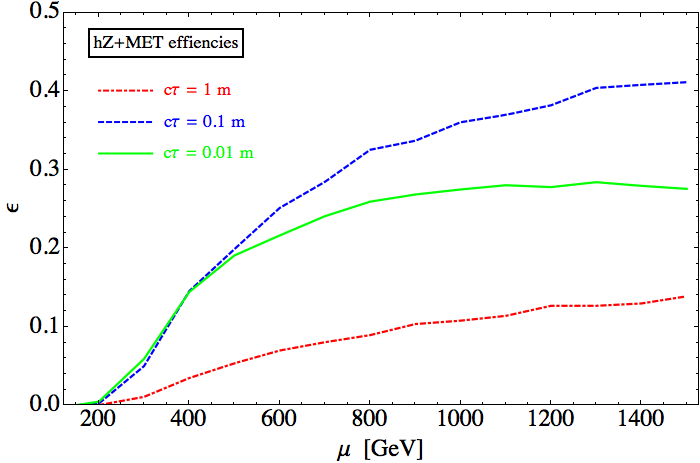} 
 \\
  \includegraphics[width=0.45\textwidth]{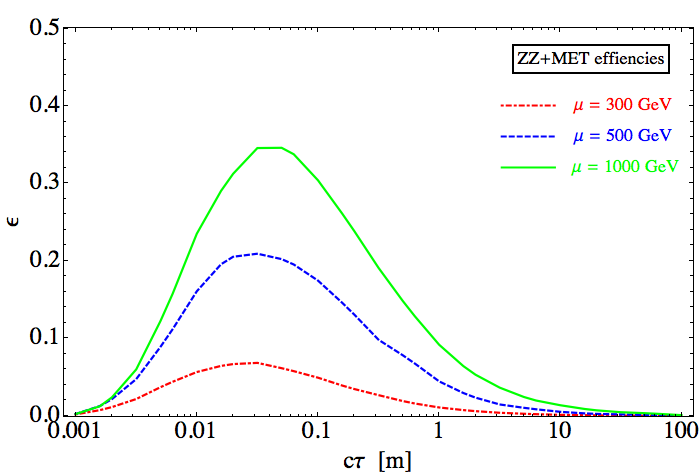}\hspace{5mm}
 \includegraphics[width=0.45\textwidth]{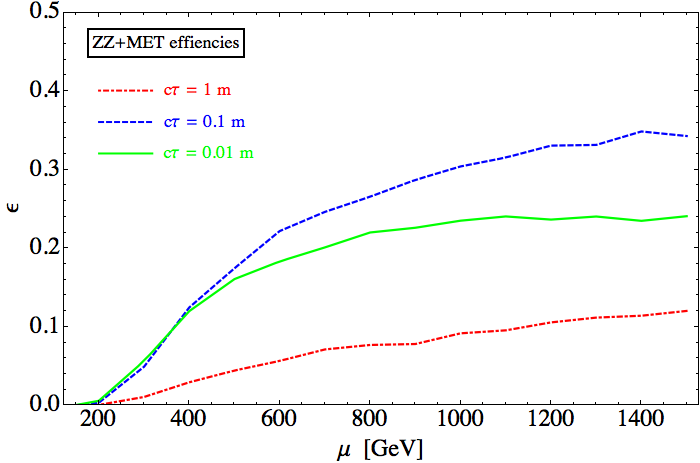} 
 \caption{\label{fig:eff_model}
Efficiencies for the simplified final states with $hh+\MET$, $hZ+\MET$, $ZZ+\MET$. Left: efficiencies as a function of $c \tau$ for $m_{\chi_2}=m_{\chi_3}=300,\,500,\,1000$ GeV.
Right: efficiencies as a function of the mass $m_{\chi_2}=m_{\chi_3}$ for $c\tau=1,\, 10,\, 100$ cm.
 }
\end{figure}

In Figure \ref{fig:valid} we show our estimated cross section exclusion compared to the ATLAS results. 
We find a good agreement in the region of un-compressed spectrum, while our simulation overestimate the exclusion power 
in the compressed spectrum case. We argue that this is due to our implementation of the jet cut (number 2.~in the list above), for which we have only a limited amount of information provided by the ATLAS documentation.
The plots of Figure \ref{fig:valid} shows that, on the other hand, our simulation consistently reproduces the ATLAS analysis for a mass difference between the heavy particle and its decay products larger than $\approx100$ GeV, taking into account an uncertainty of $\pm 50\%$ on the efficiency.

In Figure \ref{fig:eff_model} we display the result of our recasting: the efficiency curves for the final states characterising the Singlet-Doublet model. 
The samples have been generated at LO with {\tt MadGraph5} + {\tt Pythia3} + {\tt Delphes3} after implementing the model in {\tt FeynRules}.
We simulated the following three cases
$$
p p \to \chi_2 \chi_3 \to ZZ \chi_1 \chi_1, \qquad p p \to \chi_2 \chi_3 \to hh \chi_1 \chi_1, \qquad p p \to \chi_2 \chi_3 \to hZ \chi_1 \chi_1,
$$
where the decay of the bosons is performed in {\tt Pythia}.
We then processed the output with the selection procedures explained above to extract the efficiencies as a function of the mean lifetime $c\tau$ and the mass of the decaying heavy particles $m_{\chi_2}=m_{\chi_3}$.

As a final remark, let us notice that the efficiencies displayed in Figure \ref{fig:eff_model} have been obtained by treating displaced heavy flavour jets like light-flavour ones.
However, the case of $Z$ or $h$ decaying into $b\bar{b}$ pairs requires in principle additional care, since the recast DV+$\MET$ search associates tracks to a displaced vertex based on track-vertex compatibility requirements, and merges displaced vertices if within 1 mm. For displaced $b$ jets, these requirements are difficult to recast. As we mentioned, we choose to neglect this possible issue (an interesting discussion of which can be found
in Ref.~\cite{Allanach:2016pam}). 
We argue that this simplification has a limited impact on our estimated exclusions (shown in Figures~\ref{fig:LHC_bound_yplot} and \ref{fig:combi_plot}) for the following reasons: 
(i) Only one DV is enough to satisfy the analysis' requirements, thus there is no loss of sensitivity if at least one of the pair-produced heavy particles decays into a $Z$ decaying into light flavours (and the DV is reconstructed);
(ii) Due to the gluons radiated by the $b$ quarks, a DV can still be formed on the tracks not coming from the $b$-decay vertex;
(iii) Part of the $b$ decays will still happen within the required 1 mm.
%
%
\newpage
\bibliographystyle{JHEPours} 
\bibliography{bibSD-FI}
%
%
\end{document}